\documentclass[twocolumn,showpacs,preprintnumber,amsmath,amssymb,aps]{revtex4}
\usepackage{graphicx}
\begin{document}

\title{  
Transport Anisotropy as a Probe of the Interstitial Vortex State in Superconductors with
Artificial Pinning Arrays   
} 
\author{C. Reichhardt and C.J. Olson Reichhardt} 
\affiliation{ 
Theoretical Division, 
Los Alamos National Laboratory, Los Alamos, New Mexico 87545}

\date{\today}
\begin{abstract}
We show using simulations 
that when interstitial vortices are present in superconductors
with periodic pinning arrays, the transport in two perpendicular 
directions can be anisotropic.
The degree of the anisotropy varies as a function of field 
due to the fact that the interstitial vortex lattice has 
distinct orderings at different matching fields.  
The anisotropy is most pronounced at the matching fields but 
persists at incommensurate fields, and it is
most prominent for triangular, honeycomb, and kagom{\' e} pinning arrays.  
Square pinning arrays can also show anisotropic
transport at certain fields in spite of the fact that
the perpendicular directions of the square pinning array are identical.
We show that the anisotropy results from 
distinct vortex dynamical states and that although
the critical depinning force may be lower in one direction, 
the vortex velocity above depinning may also be lower in the same
direction 
for ranges of external drives where both directions are depinned. 
For honeycomb and kagom{\' e} pinning arrays, 
the anisotropy can show multiple reversals as a function of field.    
We argue that when the pinning sites can be multiply occupied such that
no interstitial vortices are present, the anisotropy is strongly
reduced or absent.
\end{abstract}
\maketitle

\vskip2pc

\section{Introduction}

Vortices in superconductors interacting with 
artificial arrays of periodic pinning exhibit 
a wide range of  
commensurability and 
dynamical effects that can be observed 
readily in critical current, transport, and other bulk 
measurements 
\cite{Fiory,Metlushko,Baert,Pannetier,Rosseel,Harada,Welp,Kwok,Martin,Ketterson}. 
Advances in lithography techniques permit the creation of 
pinning arrays in which the size, shape, and composition of the
individual pinning sites and the global geometry can be well controlled 
\cite{Metlushko,Baert,Pannetier,Rosseel,Harada,Welp,Kwok,Martin,Ketterson,Zhukov,Raedts,Goran,Fasano}.  
Commensurability effects in these systems
occur when the number of vortices equals an integer 
multiple of the number of pinning sites,
resulting in peaks or anomalies in bulk measurements as  
a function of field. 
At the first matching field, there is one vortex per pinning site, and
as the field is further increased, 
additional vortices can be located either at the pinning sites 
in the form of pinned multi-quanta vortices
\cite{Baert,Pannetier,Field,Bending,Jensen,Schuller,Peeters}, or 
in the interstitial regions between the pinning sites. 
The interstitial vortices can 
be effectively pinned by the repulsive interactions 
from the vortices at the pinning sites, which 
create a caging potential 
\cite{Harada,Rosseel,Reichhardt,Chen,Milosvic,Peeters,Zimanyi1,Olson}. 
It is also possible for mixed vortex pinning to occur in which
the first few matching fields have only pinned multi-quanta vortices
until the pinning sites are saturated, while for higher matching fields the
additional vortices are located in the interstitial regions
\cite{Baert,Harada,Goran,Field,Bending,Peeters}. 
Conversely, it is also possible that interstitial vortices appear at the
lower matching fields, but that as the vortex-vortex interactions increase
at higher matching fields, multi-quanta vortices will begin to form at the
pinning sites \cite{Peeters}.

Interstitial vortex lattice crystals in square periodic pinning arrays 
have been observed directly with Lorentz microscopy, 
which revealed that there are several distinct
types of interstitial vortex structures that have 
symmetries different from that of the triangular vortex lattice \cite{Harada}.
The same types of vortex structures have been produced in simulations 
of square pinning arrays \cite{Reichhardt,Chen}, 
while simulations have also shown that similar vortex structures can form
in triangular \cite{Reichhardt}, rectangular \cite{Zimanyi,Olson}, 
honeycomb \cite{Molecular}, and kagom{\' e} pinning 
arrays \cite{Molecular,Dom}.  
Other numerical works indicated 
that a rich variety of composite lattices with multiple and interstitial 
vortex  
configurations are possible \cite{Milosvic,Peeters} and that
new types of interstitial vortex configurations can occur 
for arrays of antipinning sites \cite{Peet}.  

Vortex imaging experiments provide direct 
evidence for both multi-quanta vortex pinning and 
the formation of ordered interstitial
vortex lattice 
structures \cite{Field,Bending}.  
Anomalies at matching fields found in bulk measurements occur for both
multi-quanta vortex pinning and interstitial vortex pinning,
so without direct imaging it can be difficult to determine
whether multi-quanta or interstitial vortex pinning is 
occurring \cite{Schuller}. In some cases, 
the presence of interstitial vortices
can only be inferred from the
shapes and characteristics of the current voltage curves or 
from phase locking experiments \cite{Rosseel,Shapirolook}. 
It would be highly desirable to identify additional clear 
signatures in transport measurements
that can distinguish between interstitial vortex pinning 
and multi-quanta vortex pinning
and that can also reveal the types of vortex 
lattice symmetries that are present.

An anisotropic response was recently measured for the critical current
applied in two perpendicular directions to a triangular  pinning
array in recent experiments and simulations \cite{Wu,Cao}.
The experiments were performed 
on several different samples and the same anisotropic response appeared
in each one, while the anisotropy observed in the simulations agreed with
that seen in the experiments, 
suggesting that 
the behavior is due to distinct intrinsic features of the vortex dynamics. 
The simulations show that the vortex flow patterns 
are different for the two directions of applied current,
which could account for the anisotropic response.  
The anisotropy is particularly pronounced at the
second matching field but is absent at the 
first matching field, which suggests that 
depinning of the interstitial vortices is responsible for the anisotropy. 
Interestingly, for lower temperatures the experiments showed that the
critical current anisotropy vanished at both the first and 
second matching fields but persisted at the third matching
field, suggesting that multi-quanta vortex pinning 
occurs at the second matching field at low temperatures. 
These results indicate that the presence of anisotropy 
can be a useful way to probe the interstitial vortex state and the 
type of vortex ordering that occurs at different matching fields. 

The experiments and simulations of Refs.~\cite{Wu,Cao} only examined a
triangular pinning array up to the third matching field.
In this work we study anisotropic transport for a much wider range of fields 
and system parameters for triangular, square, honeycomb, and kagom{\' e} 
pinning arrays. For the
triangular array we find a variety of distinct anisotropic behaviors 
due to the differing symmetries of the interstitial vortex lattice at different 
matching fields. For example, at certain matching fields 
the vortex lattice is disordered and the anisotropy vanishes.
For honeycomb and kagom{\' e} pinning arrays,
an even richer anisotropic behavior occurs due to the formation 
of vortex molecular crystal states \cite{Molecular} 
with additional rotational degrees of freedom, resulting in
a series of reversals in the anisotropy as a function of field. 
Remarkably, we find that it is also possible for square pinning 
arrays to show anisotropic transport even though the two perpendicular
directions of the pinning lattice are identical. 
This occurs at certain matching fields
where a triangular vortex structure forms, 
such that one driving direction is oriented with the easy-shear 
direction of the vortex lattice.
Anisotropic transport measurements for vortices in periodic pinning 
arrays have already been been 
shown to be experimentally feasible, and  
experiments have been performed in which the current is injected in
two directions for samples with rectangular pinning arrays or asymmetric
pinning shapes, revealing anisotropic depinning thresholds
\cite{Look2,Velez,Villegas2}.
Although our work 
is focused on superconducting vortices, our results should be general to 
the class of systems of particles interacting  with
periodic substrates where both pinned and interstitial particles 
are present.  Examples of this type of system 
include colloidal particles in periodic pinning \cite{Bechinger,Tierno} 
and charged metallic balls \cite{Coupier}.

\begin{figure}
\includegraphics[width=3.5in]{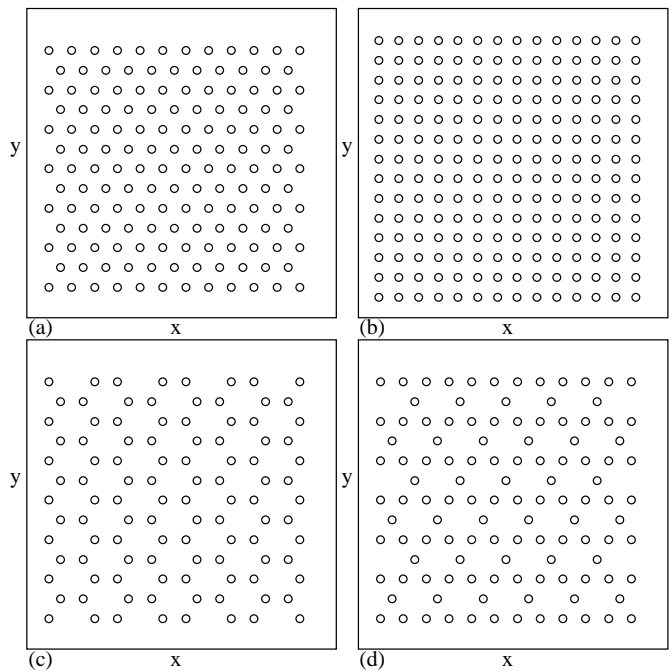}
\caption{
The pinning site locations (open circles) for 
(a) a triangular pinning array, (b) a square 
pinning array, (c) a honeycomb pinning array, and (d) 
a kagom{\' e} pinning array. 
The driving force is applied along ${\bf \hat{y}}$ to determine the
depinning force $F^{y}_c$ and along ${\bf \hat{x}}$ to determine
$F^{x}_c$.
}
\label{fig:1}
\end{figure}

\section{Simulation and System} 

We consider a two-dimensional system 
with periodic boundary conditions in the $x$ and
$y$-directions of size $L \times L$.  
The magnetic field ${\bf B}$ is applied out of the plane in the $z$-direction,  
while $N_{v}$ vortices and $N_{p}$ pinning sites are placed within
the system for a vortex density of   
$n_{v} = N_{v}/L^2$ and a pinning density of $n_{p} = N_{p}/L^{2}$. 
The matching field $B_\phi$ is defined as the 
field at which the number of vortices equals the number of pinning sites, 
$N_v=N_p$.
In Fig.~\ref{fig:1} we show representative examples of 
the triangular, square, honeycomb, and
kagom{\' e} pinning geometries used in this work, 
and indicate the ${\bf \hat{x}}$ and ${\bf \hat{y}}$ 
directions along which current is applied.
The initial vortex positions are prepared by simulated annealing
with no applied drive, and
then the vortex velocities are measured in the presence of
a driving force that is applied in the $x$ direction.
The driving force corresponds to the Lorentz force generated by an applied
current, while the vortex velocities are proportional to the voltage response
that would be measured experimentally.
We repeat the simulation from the same initial vortex positions
with the driving force applied in the
$y$ direction, and compare the velocity response 
$\langle V_x\rangle$ and $\langle V_y\rangle$ and the critical
depinning force $F^y_c$ and $F^x_c$ for the two driving directions.

The time evolution of the vortex dynamics is governed by 
integrating $N_{v}$ coupled overdamped equations of motion.
The equation of motion for a single vortex $i$ 
at position ${\bf R}_i$ is given by
\begin{equation}
\eta\frac{d {\bf R}_{i}}{dt} =  {\bf F}^{vv}_{i} + {\bf F}^{vp}_{i} +  {\bf F}_{D} + {\bf F}^{T}_{i}  .
\end{equation} 
Here the damping constant $\eta = \phi^{2}_{0}d/2\pi \xi^2 \rho_{N}$,
where $d$ is the sample thickness, $\eta$ is the coherence length,
$\rho_{N}$ is the normal-state resistivity, and $\phi_{0} = h/2e$ is the
elementary flux quantum. 
The vortex-vortex interaction force is 
\begin{equation} 
{\bf F}^{vv}_{i} = \sum^{N_{v}}_{j \neq i}f_{0}
K_{1}\left(\frac{R_{ij}}{\lambda}\right)
{\bf {\hat R}}_{ij} ,
\end{equation} 
where $K_{1}$ is the modified Bessel function, 
$\lambda$ is the London penetration depth, 
the unit of force is $f_{0} = \phi^{2}_{0}/2\pi\lambda^{3}$,
$R_{ij}=|{\bf R}_{i}-{\bf R}_j|$, and
${\bf \hat{R}}_{ij} = ({\bf R}_{i} - {\bf R}_{j})/R_{ij}$.
The vortex-vortex interaction force falls off  sufficiently 
rapidly that a cutoff  
can be placed at $R_{ij} = 6\lambda$.
Use of a longer cutoff of $R_{ij}=12\lambda$ produces identical results.
An additional short range cutoff is placed at $R_{ij}=0.1\lambda$
to avoid a divergence in the force.
The pinning sites are modeled as attractive parabolic wells of radius $R_p$ 
and strength $F_p$ with
${\bf F}^{vp}_{i} = \sum_k^{N_p}f_0F_{p}R_{p}^{-1}R_{ik}^{(p)}
\Theta((R_{p} - R_{ik}^{(p)})/\lambda){\bf {\hat R}}^{(p)}_{ik}$.
Here ${\bf R}^{(p)}_{k}$ is the location of pinning site $k$, 
$R_{ik}=|{\bf R}_i-{\bf R}_k^{(p)}|$,
${\bf {\hat R}}_{ik}^{(p)}=({\bf R}_i-{\bf R}_k^{(p)})/R_{ik}$,
and
$\Theta$ is the Heaviside step function. 
The pinning sites are arranged in a triangular, square, 
honeycomb, or kagom{\' e} array. 
The external drive ${\bf F}_{D}=F_D{\bf {\hat x}}$ or
${\bf F}_{D}=F_D{\bf {\hat y}}$ is a constant force that is 
uniformly applied to all of the vortices.
The thermal force $F^{T}_{i}$ is used during the simulated annealing procedure
and has the following properties:
$\langle F^{T}_{i}(t)\rangle = 0$ and 
$\langle F^{T}_{i}(t)F^{T}_{j}(t^{\prime})\rangle = 
2\eta k_{B} T\delta_{ij}\delta(t- t^{\prime})$.    
We decrement the temperature
by $0.0002$ every 1000 simulation time steps. 
After the initialization, the applied drive is imposed 
in increments of $\delta F_{D} = 0.1$ every $10^3$ 
simulation time steps. 
The velocity-force curves are obtained by averaging 
the velocity every $10^3$ simulation time steps:     
$\langle V_{\alpha}\rangle = N_{v}^{-1}\sum^{{N}_{v}}_{i} {\bf v}_{i}\cdot {\bf {\hat \alpha}}$, where $\alpha=x$, $y$.
Here ${\bf v}_{i} = d{\bf R}_{i}/dt$.   
The critical depinning forces in the $x$ and $y$ directions,
$F_c^x$ and $F_c^y$, 
are determined by the criterion $\langle V_\alpha\rangle=0.001$.

\begin{figure}
\includegraphics[width=3.5in]{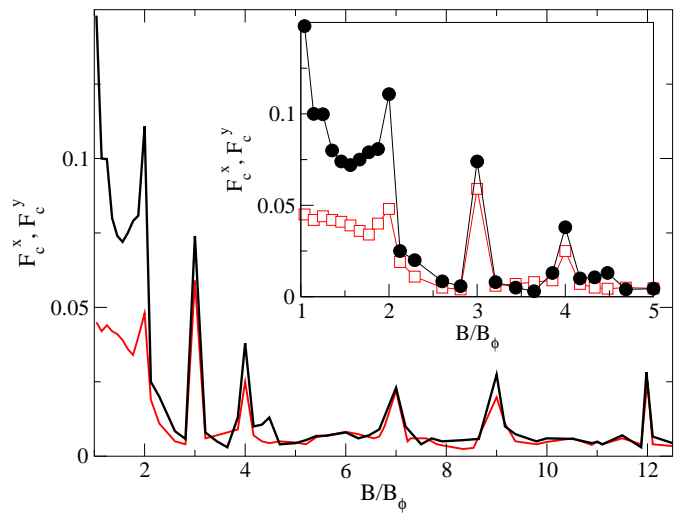}
\caption{
The depinning forces $F^{x}_{c}$ (light line) and $F^{y}_{c}$ (dark lines) 
vs $B/B_{\phi}$ for 
a triangular pinning lattice with $F_{p} = 0.85$, 
$R_{p} = 0.35\lambda$, and $n_p = 0.0833/\lambda^2$.
Inset: the same data showing a 
highlight of the region from $1.0 < B/B_{\phi} < 5.0$. 
Open squares: $F^x_c$; filled circles: $F^y_c$.
}
\label{fig:2}
\end{figure}

\begin{figure}
\includegraphics[width=3.5in]{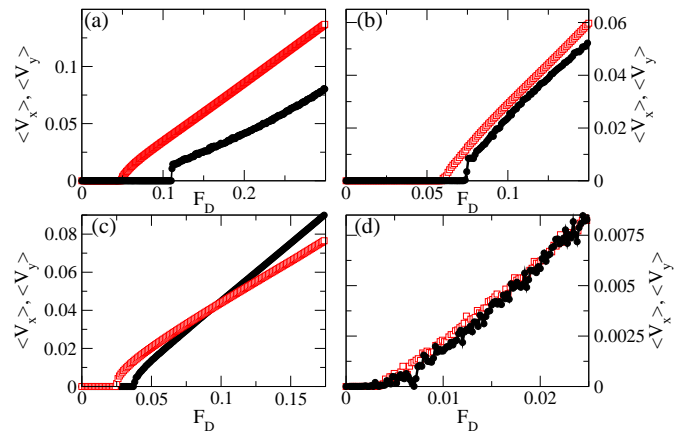}
\caption{ 
Velocity-force curves for driving in the $x$-direction,
$\langle V_x\rangle$ (open squares), and 
in the $y$-direction, $\langle V_y\rangle$ (filled circles) 
for the triangular pinning
lattice system in Fig.~\ref{fig:2}.   
(a) At $B/B_{\phi} = 2$, the slope $dV_y/dF_D < dV_x/dF_D$. 
(b) At $B/B_{\phi} = 3$, 
$dV_y/dF_D=dV_x/dF_D$.
(c) At $B/B_{\phi} = 4$, 
$dV_x/dF_D<dV_y/dF_D$,
resulting in a crossing in the velocity-force curves near $F_{D} = 0.1$.  
(d) At $B/B_{\phi} = 5$, the depinning is isotropic.
}
\label{fig:3}
\end{figure}

\begin{figure}
\includegraphics[width=3.5in]{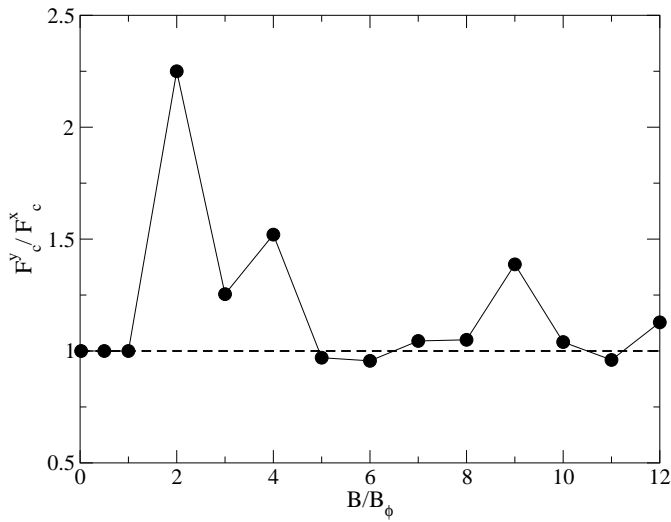}
\caption{
The ratio $F^{y}_{c}/F^{x}_{c}$ vs $B/B_{\phi}$ obtained from the system in 
Fig.~2. The dashed line indicates $F^y_c/F^x_c=1.$ 
The depinning is strongly anisotropic at the matching fields 
$B/B_{\phi} = 2$, 3, 4, 9, and 12. 
}
\label{fig:4}
\end{figure}

\section{Anisotropy in Triangular Pinning Arrays} 

We first consider the depinning forces for the $x$ and $y$-directions 
in the triangular pinning lattice
illustrated in Fig.~\ref{fig:1}(a) with $F_p=0.85$, $n_p=0.0833/\lambda^2$,
and $R_p=0.35\lambda$. 
In Fig.~\ref{fig:2} we plot 
$F_{c}^x$ and $F_c^y$ vs $B/B_{\phi}$ along with a detail of 
the region from $ 1.0 < B/B_{\phi} < 5.0$. 
In general $F_{c}^y>F_c^x$ 
at most of the commensurate fields.
The anisotropy at incommensurate fields is most pronounced
for $1.0 < B/B_\phi < 3.0$, as shown in 
the inset of Fig.~\ref{fig:2}.
In previous simulations a similar anisotropy 
was observed for this range of fields \cite{Cao}. For $B/B_{\phi} > 3.0$
both $F_c^x$ and $F_c^y$ are small,
so it is difficult to determine whether the depinning is anisotropic
at incommensurate fields.
At the matching 
fields $B/B_\phi=7$, 9, and 12, where the depinning forces are high, 
$F_{c}^x$ and $F_c^y$ can be measured accurately.
In Fig.~\ref{fig:3} we plot
representative velocity-force curves for driving in both the $x$ and
$y$ directions at $B/B_\phi=2$, 3, 4, and 5.
There is a clear anisotropy in the depinning force with $F_c^y>F_c^x$ at
$B/B_\phi=2$, 3, and 4 in Figs.~\ref{fig:3}(a,b,c), 
while at $B/B_{\phi} = 5$ in Fig.~\ref{fig:3}(d),
the depinning is isotropic.
This can be seen more clearly in the plot of 
the anisotropy $F^y_c/F^x_c$ at the
matching fields, shown in Fig.~\ref{fig:4}.
The anisotropy is largest for $B/B_{\phi} = 2,$ 4, 
and 9, weaker for $B/B_\phi=3$ and 12, and essentially absent at
$B/B_\phi=1$, 5, 6, 7, 8, 10, and 11. 
Corresponding to this, there are no peaks in the depinning force at
$B/B_\phi=5$, 6, 8, 10, and 11 in Fig.~\ref{fig:2}.
At the fields with isotropic depinning $F_c^y/F_c^x \approx 1$, 
the overall vortex lattice is disordered, while at the other matching 
fields where anisotropic depinning occurs, ordered vortex lattices form.
Previous numerical work for triangular pinning arrays at fields up 
to $B/B_{\phi}=9$ showed the same features 
in the critical depinning force as well as 
the existence of disordered lattices at the matching fields 
$B/B_\phi=5$, 6, and 8 \cite{Reichhardt}. 
From a geometric construction, it can be shown 
that a triangular vortex lattice can be placed on a triangular pinning 
lattice at the integer matching 
fields $N =  m^2 + n^2 + nm$, where $n$ and $m$ are integers
\cite{Reichhardt}. This predicts
the formation of triangular vortex lattices at fields with $N = 1$, 3, 4, 7, 
9, and 12, in agreement with our observation of a peak in $F_c$ at each of 
these fields. 
The geometric construction
does not predict the formation of 
a triangular lattice at $B/B_{\phi} = 2$; however, a strong matching 
peak appears at this field both in Fig.~\ref{fig:2} 
and in the previous work \cite{Reichhardt}. 
The peak at $B/B_\phi=2$ occurs due to the formation of an
ordered honeycomb vortex lattice structure, rather than a triangular vortex
lattice.
In general, we expect to find a peak in the critical
current at fields where a triangular or other 
ordered vortex lattice structure forms. 
Figures~\ref{fig:2} and \ref{fig:4} also show that although there is a peak
in $F_c$ at $B/B_{\phi} = 7$, where a triangular vortex lattice
forms, there is no anisotropy for this field and $F_c^y/F_c^x\approx 1$. 

\begin{figure}
\includegraphics[width=3.5in]{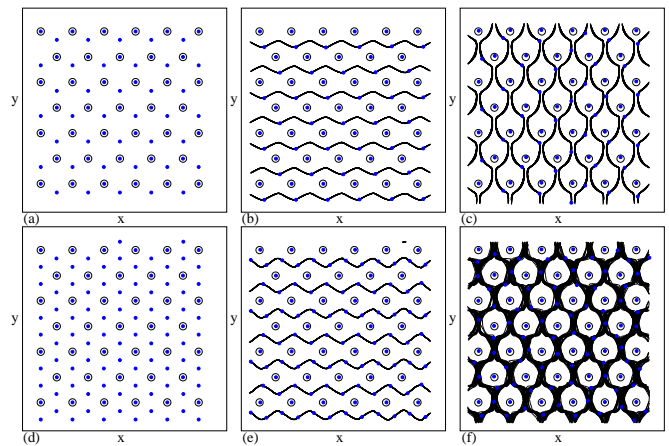}
\caption{Vortex positions (black dots), 
pinning site locations (open circles), and vortex trajectories (black lines) 
for the triangular pinning lattice system in Fig.~2. 
(a) The vortices form a honeycomb configuration 
at $F_{D} = 0$ and $B/B_{\phi} = 2$. 
(b) The vortex trajectories just above depinning 
for driving in the $x$-direction at $B/B_\phi=2$.  The vortices
channel between the pinning rows. 
(c) Vortex trajectories just above depinning for driving in the 
$y$-direction at $B/B_\phi=2$.  The moving 
interstitial vortices wind around the occupied pinning sites.
(d) The vortices form a triangular configuration at 
$F_{D} = 0$ and $B/B_{\phi} = 3$.
(e) The vortex trajectories just above depinning 
for driving in the $x$-direction at $B/B_\phi=3$ show an 
ordered interstitial flow of vortices
between the pinning sites. 
(f) Vortex trajectories just above depinning for driving 
in the $y$-direction at $B/B_\phi=3$.  Here the flow 
is disordered.
}
\label{fig:5}
\end{figure}

In order to explain the different degrees of anisotropy that appear at
different fields,
we analyze the vortex positions and trajectories
for driving in the $x$ and $y$ directions. 
Figure \ref{fig:5}(a) shows the ordered honeycomb vortex lattice structure
that forms at $F_D=0$ for $B/B_\phi=2$ on a triangular pinning lattice.
When driven in the $x$-direction, the vortices can easily form slightly
undulating channels of flow that pass between the filled pinning sites, 
as illustrated in Fig.~\ref{fig:5}(b). 
For driving in the $y$-direction, the pinned vortices act as barriers that
prevent the formation of 
simple flow channels, and the interstitial vortices can only move by making
significant excursions into the $x$-direction, as illustrated in 
Fig.~\ref{fig:5}(c) for a drive just above depinning.
As a result, 
a larger external force is required to cause depinning in the 
$y$-direction, and $F_c^y>F_c^x$. 
The onset of motion in the $y$-direction is very sharp, 
as seen by the jump in the velocity-force curve 
in Fig.~\ref{fig:3}(a), and after a brief initial period of disordered
motion, the vortex flow quickly organizes into the
pattern shown in Fig.~\ref{fig:5}(c). 
We observe similar motions just above
depinning in the $x$ and $y$-directions at 
the incommensurate fields for $1.0 < B/B_{\phi} < 2.0$; however, the 
presence of vacancies in the honeycomb vortex lattice
causes certain rows of interstitial vortices to depin at a lower value of 
$F_{D}$ than at the commensurate fields.  
The velocity-force curves at $B/B_{\phi} = 2.0$ shown in Fig.~\ref{fig:3}(a) 
indicate that $dV_y/dF_D<dV_x/dF_D$, meaning that the vortex velocity 
has a lower slope as a function of increasing drive in the $y$-direction 
than in the $x$-direction, even though the same number of vortices are 
moving for either direction of drive. This
difference is a result of the fact that the vortex motion for 
$y$-direction driving has much larger excursions transverse to the driving
direction since the interstitial vortices must go out of their way to
pass around the pinned vortices.

At $B/B_{\phi} = 3$ on a triangular pinning lattice, the overall 
vortex lattice ordering at $F_{D} = 0$ is triangular, as shown in 
Fig.~\ref{fig:5}(d). 
The vortices move in slightly winding channels upon application of a drive
in the $x$-direction, as seen in Fig.~\ref{fig:5}(e); these structures are
similar to the channels that form at $B/B_\phi=2$ in Fig.~\ref{fig:5}(b).
For depinning in the $y$-direction at $B/B_\phi=3$, 
the pinned vortices again create a barrier to the formation of
simple channels of interstitial vortex flow; however, unlike the 
ordered and strongly winding channels that form at
$B/B_\phi=2$, for $B/B_\phi=3$ the vortices move in a disordered fashion
just above depinning, as illustrated in Fig.~\ref{fig:5}(f).
An ordered flow state similar to that shown in Fig.~\ref{fig:5}(c) 
does not occur for $B/B_\phi=3$ until a much higher value of $F_{D}$ is applied.
This result indicates that in addition to the anisotropy 
in $F^y_c/F^x_c$, there are also strongly different
vortex velocity fluctuation characteristics 
for driving along the $x$ and $y$-directions at $B/B_{\phi} = 3$.
Narrow band noise signatures should arise from the synchronized vortex
motion that occurs for driving along the $x$-direction,
while for 
driving in the $y$-direction the velocities are more random 
and a broad band noise signature should appear.
In mode-locking experiments, where an external ac drive is imposed 
along with an applied dc drive, Shapiro type steps would appear for the
ordered motion along the $x$-direction, while Shapiro steps would be 
absent for driving along the $y$-direction. 
Shapiro steps could be induced for both
driving directions at $B/B_{\phi} = 2$ since both vortex flow patterns
show synchronized motion; however, some of the characteristics of the Shapiro 
steps might differ for the two directions since the meandering of the 
vortices is distinct in the $x$ and $y$-directions.   

\begin{figure}
\includegraphics[width=3.5in]{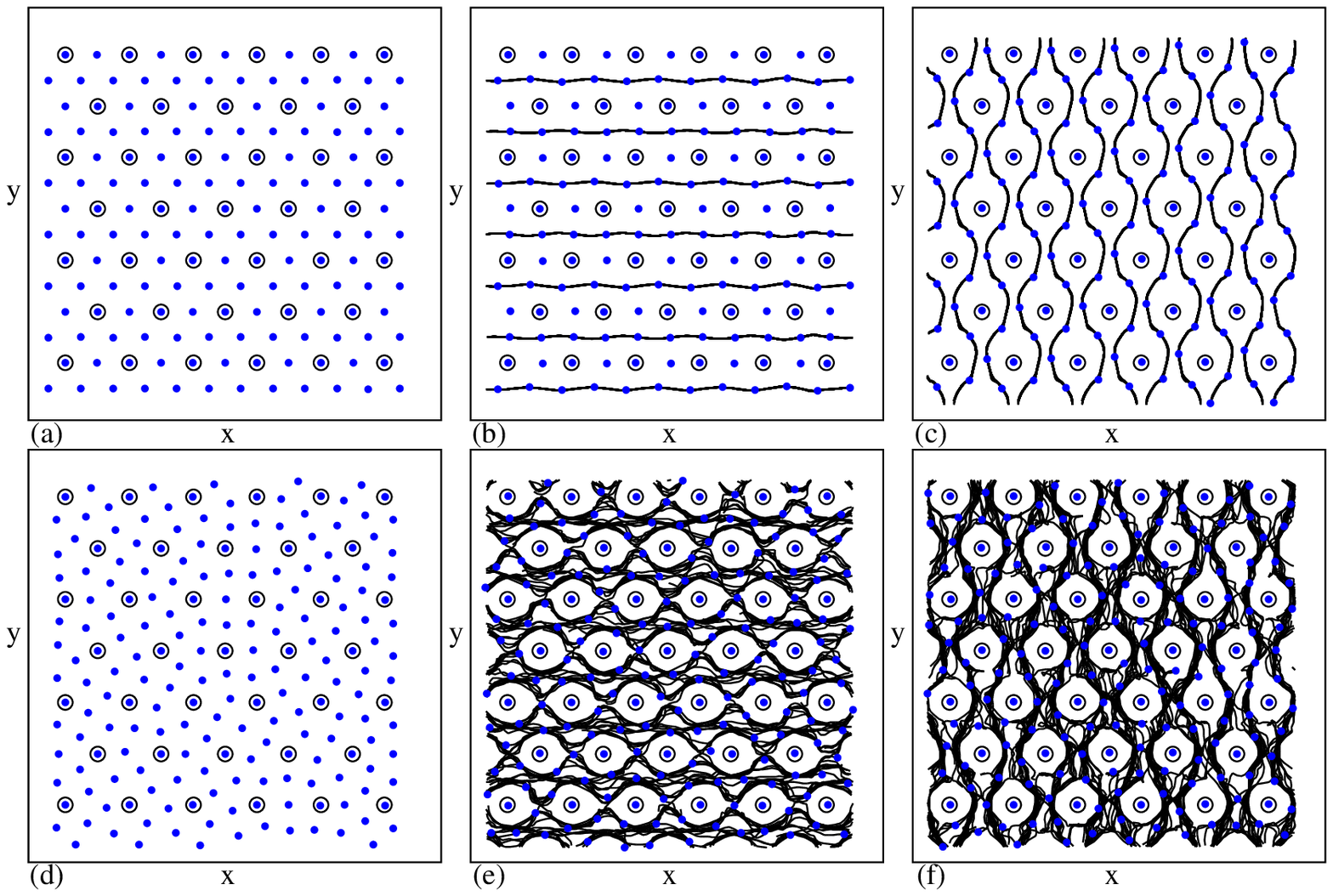}
\caption{
Vortex positions (black dots), pinning site locations (open circles), 
and vortex trajectories (black lines) for the triangular pinning 
lattice system in
Fig.~\ref{fig:2}.
(a) At $F_{D} = 0$ and $B/B_{\phi} = 4$ a triangular
vortex lattice forms. (b) The vortex trajectories 
just above depinning for driving in the
$x$-direction at $B/B_\phi=4$.
A portion of the interstitial vortices move in 
one-dimensional channels between the rows 
of pinning sites. 
(c) The vortex trajectories just above depinning 
for driving in the $y$-direction at $B/B_\phi=4$.  The interstitial vortices
flow in winding channels around the pinning sites. 
(d) At $F_{D} = 0$ and $B/B_{\phi} = 5$, 
the vortex lattice is disordered. 
(e) The vortex trajectories just above depinning
for driving in the $x$-direction at $B/B_\phi=5$.
The vortex flow pattern is disordered. 
(f) Vortex trajectories just above depinning
for driving in the $y$-direction at $B/B_\phi=5$. 
The same type of disordered flow pattern seen for $x$-direction driving
appears for $y$-direction driving.           
}
\label{fig:6}
\end{figure}

A triangular vortex 
lattice with a single row of interstitial vortices between each
row of pinning sites forms for $B/B_{\phi} = 4$ at $F_{D} = 0$, 
as shown in Fig.~\ref{fig:6}(a).
Just above the depinning transition for driving in the $x$-direction, 
a portion of the interstitial vortices depin into the one-dimensional channels 
illustrated in Fig.~\ref{fig:6}(b).
Here, one-third of the interstitial vortices remain 
immobile in the interstitial regions between pinning sites; 
these immobile interstitial 
vortices depin at a higher value of $F_{D}$ that is outside
the range of driving forces shown in Fig.~\ref{fig:3}(c). 
For driving in the $y$-direction,
all of the interstitial vortices depin simultaneously and
flow in winding channels around the occupied pinning sites, 
as shown in Fig.~\ref{fig:6}(c). 
The velocity-force curves shown in Fig.~\ref{fig:3}(c) 
indicate that although $F_c^y>F_c^x$ for $B/B_{\phi} = 4$, 
$dV_x/dF_D<dV_y/dF_D$.  As a result,
the two velocity-force curves cross near $F_{D} = 0.1$.
The slope $dV/dF_D$ is steeper for the 
$y$-direction driving 
since all the interstitial vortices take part in the motion, 
whereas 
for the $x$-direction driving, 
only $2/3$ of the interstitial vortices are moving.    

At $B/B_{\phi} = 5$ and $F_D=0$ the vortex lattice is disordered, 
as seen in Fig.~\ref{fig:6}(d), and
the depinning is isotropic.
Figure 6(d,e) shows that the same type of disordered vortex motion 
occurs for depinning in both the $x$ and $y$-directions. 
In general, we observe disordered flow states 
in both driving directions at the other matching fields where  
a disordered  vortex lattice forms, including $B/B_{\phi} =  6$, 8, 10,
and 11. 

\begin{figure}
\includegraphics[width=3.5in]{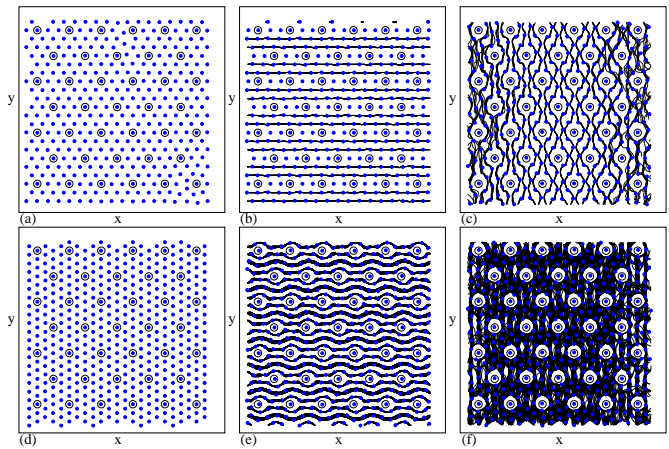}
\caption{
Vortex positions (black dots), pinning site locations (open circles), 
and vortex trajectories (black lines) for the triangular pinning
lattice system in Fig.~2. 
(a) At $F_{D} = 0$ and $B/B_{\phi} = 9$,  a triangular vortex lattice forms.
(b) The vortex trajectories just above depinning for driving
in the $x$-direction at $B/B_\phi=9$. Two rows of 
interstitial vortices move in one-dimensional paths between the 
rows of pinning sites while a portion of the interstitial
vortices remain pinned. 
(c) The vortex trajectories just above depinning for driving in 
the $y$-direction at $B/B_\phi=9$.  A combination 
of ordered and disordered flow occurs. 
(d) At $F_{D} = 0$ and $B/B_{\phi} = 12$, a triangular vortex lattice forms
which is aligned with the $y$ direction.
(e) The vortex trajectories just above depinning 
for driving in the $x$-direction at $B/B_\phi=12$.  Three rows of interstitial
vortices move between the rows of pinning sites. 
(f) Vortex trajectories just above depinning for driving 
in the $y$-direction at $B/B_\phi=12$ show 
the existence of disordered flow.
}
\label{fig:7}
\end{figure}

At  $B/B_{\phi} = 9$, the depinning thresholds and 
velocity-force curves are very similar to those 
found for $B/B_{\phi} = 4$ since the triangular vortex lattices that
form at these two fields have the same orientation.
In Fig.~\ref{fig:7}(a), the vortex configuration at $F_D=0$ for
$B/B_\phi=9$ contains two rows of interstitial vortices between adjacent
pairs of pinning rows, whereas at $B/B_\phi=4$, Fig.~\ref{fig:6}(a)
shows that there is only one row of
interstitial vortices between each pair of pinning rows.
Just above the depinning transition for $x$-direction driving at $B/B_\phi=9$,
Fig.~\ref{fig:7}(b) indicates that
the two rows of interstitial vortices flow in one-dimensional channels
between the pinning rows while two interstitial vortices remain trapped
behind every pinning site so that $3/4$ of the interstitial vortices are
moving.
This is similar to the $x$-direction depinning that occurs
for $B/B_{\phi} = 4$, where a single row of interstitial vortices flows between
each pair of pinning rows and a single interstitial vortex is trapped behind
each pinning site.
Just above the depinning transition for $y$-direction 
driving at $B/B_\phi=9$, shown in 
Fig.~\ref{fig:7}(c), all of the interstitial vortices are 
depinned and a combination of ordered and disordered vortex flow occurs.
The velocity-force curves for driving in the $x$ and $y$ directions show
a similar crossing at $B/B_\phi=9$ as that illustrated in Fig.~\ref{fig:3}(c)
for $B/B_\phi=4$.

For  $B/B_{\phi} = 12$, 
the vortex configurations and depinning dynamics are similar to those seen 
for $B/B_{\phi} = 3$. In Fig.~\ref{fig:7}(d) the $F_D=0$ vortex configuration
at $B/B_{\phi} = 12$ consists of a triangular lattice that is aligned 
with the $y$-direction. 
Since there is now an interstitial column of vortices that could depin and 
flow between adjacent columns of pinning sites, while there are no straight
rows of interstitial vortices, it might be expected that $F_c^y<F_c^x$ for
this field.  Instead, Fig.~\ref{fig:4} shows that $F_c^y/F_c^x \approx 1.13$,
so the depinning is still easier in the $x$-direction than in
the $y$-direction; however, the depinning anisotropy is much smaller than 
that which appears at $B/B_\phi=3$.
Vortex motion just above depinning at $B/B_\phi=12$ for driving
in the $x$-direction occurs in the form of three ordered 
winding rows passing between each pair of pinning rows,
as shown in Fig.~\ref{fig:7}(e). 
This is similar to the motion at $B/B_{\phi} = 3$ 
shown in Fig.~\ref{fig:5}(e) where one winding row of interstitial
vortices moves between the pinning rows.
For depinning in the $y$-direction at $B/B_\phi=12$,
Fig.~\ref{fig:7}(f) shows that the vortex trajectories are disordered 
in a manner similar to that found for $y$-direction 
depinning at $B/B_{\phi} = 3$, as seen in Fig.~\ref{fig:5}(f).            

\begin{figure}
\includegraphics[width=3.5in]{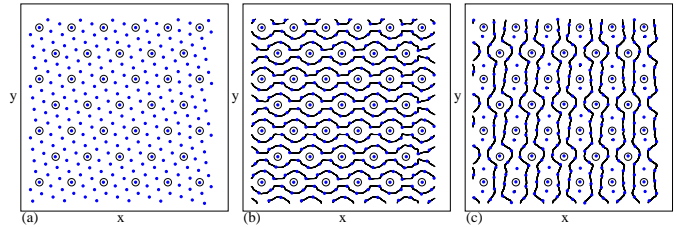}
\caption{
Vortex positions (black dots), pining site locations (open circles), 
and vortex trajectories (black lines)
for the triangular pinning lattice system in Fig.~2 at $B/B_\phi=7$. 
(a) At $F_{D} = 0$, a triangular vortex lattice forms
that is not aligned with the $x$ or $y$ axes but is tilted at an angle
$\theta \approx -78^\circ$ to the $x$-axis. 
This ground state is two-fold degenerate since the
vortex lattice could have been tilted at the opposite angle,
$\theta \approx +78^\circ$, to the $x$-axis. 
(b) The vortex trajectories just above depinning 
for driving in the $x$-direction.  All of the interstitial
vortices are moving.
(c) The vortex trajectories just above depinning for driving 
in the $y$-direction. In this case
a portion of the interstitial vortices remains pinned. 
}
\label{fig:8}
\end{figure}

At $B/B_{\phi} = 7$, a peak in $F_{c}$ occurs as shown in Fig.~\ref{fig:2},  
but there is almost no anisotropy in the depinning thresholds, as seen in 
Fig.~\ref{fig:4}. 
Figure~\ref{fig:8}(a) illustrates that at this field for $F_{d} = 0$, 
a triangular vortex lattice forms which is not aligned with 
either the $x$ or $y$ directions,
unlike the configurations found at $B/B_{\phi} = 3,$ 4, 9, and $12$, but
is at an angle $\theta \approx -78^\circ$ to the $x$-axis. 
The vortices flow in ordered patterns just above depinning in both the
$x$ and $y$ directions, as shown in Fig.~\ref{fig:8}(b,c).
Since a portion of the interstitial vortices remain pinned just above depinning
for driving in the $y$-direction, whereas all of the interstitial vortices
are flowing for driving in the $x$-direction, $dV_y/dF_D<dV_x/dF_D$ at
$B/B_\phi=7$.
The absence of the anisotropy at this field is likely 
due to the fact that the main symmetry axis of the triangular vortex 
lattice is not aligned
with either the $x$ or $y$ directions, as is the case at the other 
matching fields which show anisotropy. 

\begin{figure}
\includegraphics[width=3.5in]{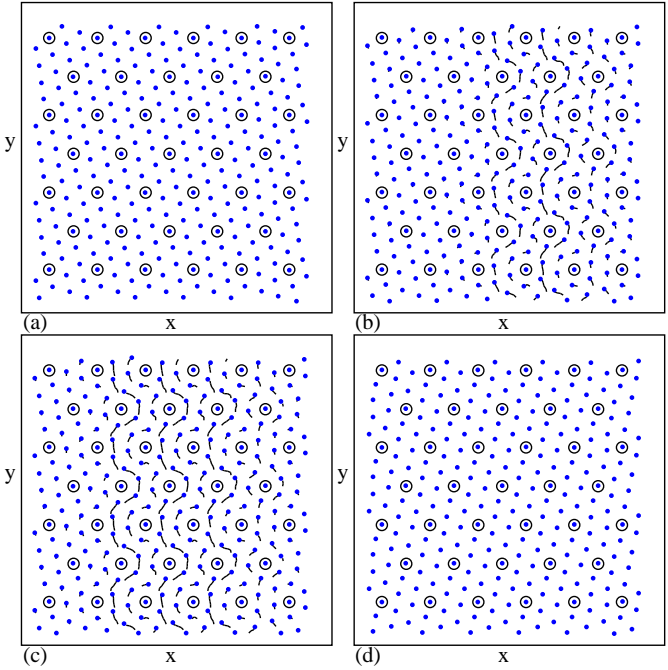}
\caption{
Vortex positions (black dots), pinning site locations (open circles),
and vortex trajectories (black lines) for the
triangular pinning lattice system in Fig.~2 at $B/B_\phi=7$. 
A weak external driving force is applied at an angle in the positive 
$x$-$y$ plane that
induces a structural transformation in the vortex lattice between the two
ground states, $-78^\circ$ shown in panel (a) and 
$+78^\circ$ shown in panel (d). 
During the transformation, illustrated in panels (b) and (c), 
the interstitial vortices shift in the positive $y$-direction by 
approximately one lattice constant and a domain wall traverses the system. 
}
\label{fig:9}
\end{figure}

An interesting feature that we observe at $B/B_\phi=7$ which does not occur
at the other matching fields we have investigated is a switching dynamics
that can be induced within the pinned phase.
This is illustrated in Fig.~\ref{fig:9} 
where in the initial state, shown in Fig.~\ref{fig:9}(a), the vortex lattice is 
tilted at $\theta \approx -78^\circ$ 
to the $x$-axis.
The ground state is twofold 
degenerate since it could also have been aligned at $\theta \approx +78^\circ$
to the $x$-axis.
When a small driving force is applied to the ground state shown in Fig.~\ref{fig:9}(a)
along $+45^\circ$ to the $x$-axis,
a structural rearrangement of the vortices occurs accompanied by a domain
wall that passes through the system from right to left, as
shown in Fig.~\ref{fig:9}(b,c). 
The vortices move by approximately one lattice constant in the $y$-direction
as the domain wall sweeps past, and in the final state the vortex lattice
is tilted at $\theta \approx +78^\circ$ to the $x$-axis.
The system can be switched back to $-78^\circ$ 
if an external drive is applied along $-45^\circ$ to the $x$-axis.
We expect that
for higher matching fields beyond the fields that we consider here,
a switching effect will be present for matching fields containing
two or more degenerate ground states 
where the vortex lattice could be arranged in several possible ways. 
The application of an external drive 
will lower the energy of one of the orientations.

\begin{figure}
\includegraphics[width=3.5in]{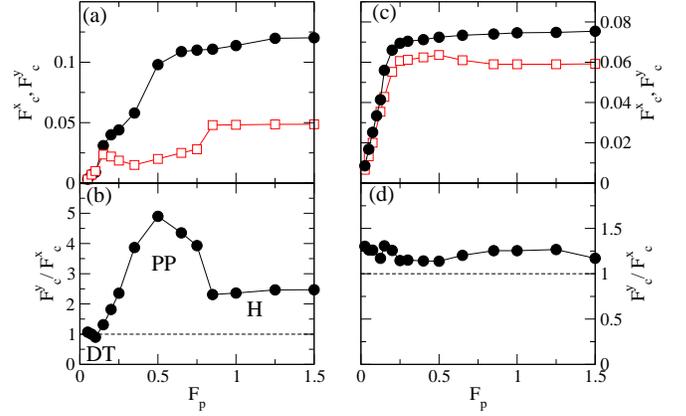}
\caption{
(a) Depinning threshold in the $x$-direction, $F^{x}_{c}$ (open squares), 
and $y$-direction, $F^{y}_{c}$ (filled circles), vs $F_{p}$ for the
triangular pinning lattice system in Fig.~2 at 
$B/B_{\phi} = 2$.  
(b) The corresponding anisotropy ratio $F^y_c/F^x_c$ vs $F_p$.
$H$: the honeycomb ordering illustrated in Fig.~\ref{fig:11}(a);
$PP$: the partially pinned phase shown in Fig.~\ref{fig:11}(b)
where a portion of the pinning sites are unoccupied 
and the depinning is plastic;
$DT$: the distorted triangular phase seen in Fig.~\ref{fig:11}(c)
where most pinning sites are unoccupied and the depinning is elastic.
The three different phases
are visible as features in the anisotropy. 
(c) The depinning thresholds $F^x_c$ and $F^y_c$ vs $F_p$ for the same system 
at $B/B_{\phi} = 3$. (d) The corresponding anisotropy ratio 
$F^y_c/F^x_c$ vs $F_p$.
}
\label{fig:10}
\end{figure}

\begin{figure}
\includegraphics[width=3.5in]{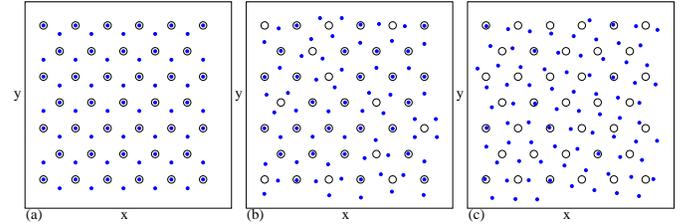}
\caption{
The vortex configurations for the triangular pinning lattice system 
at $B/B_\phi=2$ from
Fig.~10(a,b) at $F_{D} = 0$.  (a) The pinned honeycomb lattice (H)
at $F_{p} = 1.0$.
(b) The partially pinned lattice (PP) at $F_{p}  = 0.5$.  A portion of the
pinning sites are unoccupied and the vortices depin plastically. 
(c) The weak pinning regime at $F_p=0.075$ where a distorted
triangular (DT) lattice forms and the depinning is elastic.         
}
\label{fig:11}
\end{figure}
 
\subsection{Effect of Pinning Strength}

We next study the effect of varying 
the pinning strength $F_{p}$ on the anisotropy at the 
different fields. 
In Fig.~\ref{fig:10}(a) we plot $F^{y}_{c}$ and $F^{x}_{c}$ versus $F_p$ for 
a sample with a triangular pinning lattice at $B/B_{\phi} = 2$,
and in Fig.~\ref{fig:10}(b) we show the corresponding 
anisotropy ratio $F^{y}_{c}/F^{x}_{c}$.
Since the pinning induces the honeycomb vortex lattice structure shown
in Fig.~\ref{fig:11}(a) 
at this field, as $F_p$ decreases the vortices shift into configuration
that is closer to a triangular lattice.  To accommodate this shift, 
some of the pinning sites are vacated with decreasing $F_p$.
For $F_p \geq 0.85$, 
the honeycomb lattice (H) structure is stabilized,
all of the pinning sites are occupied, the anisotropy 
is fixed near $F^y_c/F^x_c=2.5$, and the
depinning thresholds $F^x_c$ and $F^y_c$ do not change 
significantly with $F_p$.
For $0.175 < F_{p} < 0.85$, the pinning is not strong enough to stabilize the
honeycomb lattice and a partially pinned (PP)
lattice forms in which only some of the
pinning sites are occupied, as illustrated in Fig.~\ref{fig:11}(b) for
$F_p=0.5$.
The depinning in this regime is still plastic with
the interstitial vortices depinning first.  
There is a strong enhancement of the anisotropy in the PP phase, 
with $F^y_c/F^x_c$ reaching values as large as 5.
For $ F_{p} \le 0.175$, the vortices form a distorted triangular (DT) 
lattice illustrated in Fig.~\ref{fig:11}(c) which becomes
increasingly triangular with decreasing $F_p$.  In the DT phase, only a small
fraction of the pinning sites are occupied.
Here the depinning transition is elastic 
and both the interstitial and pinned vortices depin simultaneously.
The anisotropy of the depinning is lost and the critical
depinning forces are isotropic in the DT phase. 
In general, $F^y_c$ and $F_c^x$ decrease with decreasing $F_p$;
however, near the transition between the PP and DT phases,
$F^{x}_{c}$ increases with decreasing $F_{p}$ as the depinning changes from
plastic to elastic. 
We believe that this effect is similar to 
the peak in the critical depinning force observed 
for periodic \cite{Zimanyi2} 
and random arrays \cite{Fertig} 
above the first matching field 
when the vortex-vortex interaction strength is 
varied. 
For periodic pinning, Ref.~\cite{Zimanyi2} illustrated that the
depinning force decreases with increasing vortex-vortex interaction
strength since the system depins elastically and all of the vortices begin
to move at the same time.
For weak vortex-vortex interactions,
the interstitial vortices 
depin easily and flow plastically past the vortices at the 
pinning sites, so the depinning force decreases with decreasing
vortex-vortex interaction strength in this regime.
Between these two extremes, a peak in the depinning force occurs. 
A similar effect is observed 
in vortex systems with very dilute random pinning arrays \cite{Fertig}. 
In Fig.~\ref{fig:10}(a) the vortex-vortex interaction
strength is fixed; however, 
there is still a transition from elastic to plastic depinning.
The peak appears only for $x$-direction driving, and this may be
due to the existence of an easy shear mode in the $x$-direction which
is absent for $y$-direction driving.

For $B/B_{\phi} = 3$, Fig.~\ref{fig:10}(c) shows that $F^{y}_{c}$ 
and $F^{x}_{c}$ saturate at large $F_{p}$, while Fig.~\ref{fig:10}(d)
indicates that the anisotropy ratio also saturates at
$ F^{y}_{c}/F^{x}_{c} =  1.24$.
For 
$F_p<0.25$, the depinning threshold decreases rapidly with decreasing $F_p$
as the system enters the elastic depinning regime.  
We note that since the vortex lattice at $B/B_\phi=3$ is triangular, 
there is no elastic energy cost for occupying the pinning sites and
one-third of the vortices will always be located at the pinning sites for
arbitrarily small $F_p$, 
unlike the situation at $B/B_\phi=2$.
There is a very small peak in $F^{x}_{c}$ near the plastic-elastic 
depinning transition.  

\begin{figure}
\includegraphics[width=3.5in]{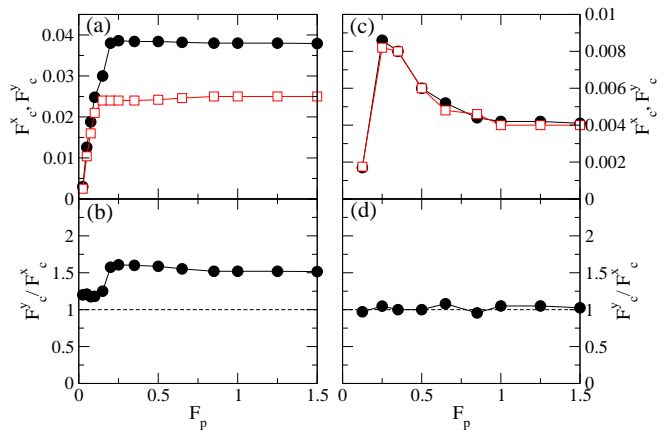}
\caption{(a) The depinning threshold in the $x$-direction, 
$F^{x}_{c}$ (open squares), and $y$-direction, $F^{y}_{c}$ (filled circles)
vs $F_p$ for the triangular pinning lattice system in Fig.~\ref{fig:2} at
$B/B_\phi=4$.
(b) The corresponding anisotropy ratio $F_c^y/F_c^x$ vs $F_p$.
(c) $F_c^x$ and $F_c^y$ vs $F_p$ for the same system at $B/B_\phi=5$.
(d) The corresponding anisotropy ratio $F_c^y/F_c^x$ vs $F_p$.
Here the depinning is isotropic 
and a peak in the depinning thresholds occurs near the transition from the
plastic to the elastic depinning regime.  
}
\label{fig:12}
\end{figure}

Figure~\ref{fig:12}(a,b) shows the depinning thresholds $F_c^x$ and $F_c^y$ as a 
function of $F_p$ for $B/B_{\phi} = 4$ along with the anisotropy
ratio $F^y_c/F^x_c$. 
Both $F_c^x$ and $F_c^y$ saturate with increasing $F_p$, and the
anisotropy for $F_p>0.25$ is fixed at $F_c^y/F_c^x=1.5$.
For $F_{p} < 0.25$,  the depinning threshold decreases rapidly with
decreasing $F_p$, and at the same time there is a
drop in the anisotropy
as the system passes from the plastic to the elastic depinning regime.
In Fig.~\ref{fig:12}(c,d) we illustrate
$F_c^x$, $F_c^y$, and $F^y_c/F^x_c$
for $B/B_{\phi} = 5$, 
where the vortex lattice lattice is disordered.
Here the depinning is isotropic for all values of $F_{p}$.   
A peak in the depinning thresholds occurs for both directions 
of driving near $F_{p} = 0.25$ at the transition between 
elastic and plastic depinning.  

We observe the same general trends for $B/B_{\phi} > 5$, including
a saturation of the anisotropy with increasing $F_p$ and
a crossover from plastic to elastic depinning at low $F_{p}$.    
For higher values of $F_{p}$ and $R_{p}$, 
multiple vortices are trapped at each pinning site. 
In this case, 
the vortices form vortex molecular crystals within the pinning sites 
\cite{Jensen2}, which changes both the
depinning threshold and the anisotropy. 
In general, the anisotropy is reduced when multiple vortex pinning is present. 
This means that the anisotropy 
is only
observable in a regime where the pinning is strong enough to allow for 
plastic depinning or channeling of vortices between the pinning sites, 
but not strong enough to permit multiple vortices to occupy each
pinning site.

\begin{figure}
\includegraphics[width=3.5in]{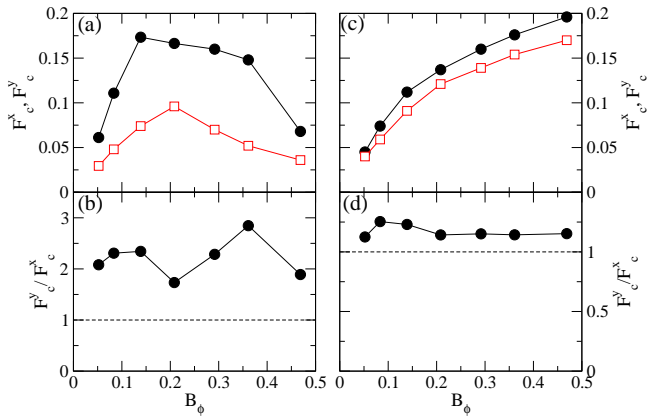}
\caption{(a) The depinning threshold in the $x$-direction, 
$F^{x}_{c}$ (open squares), and $y$-direction, $F^{y}_{c}$ (filled circles)
vs $B_{\phi}$ 
for a triangular pinning lattice system 
with $F_{p} = 0.85$ and $R_{p} = 0.35\lambda$ at $B/B_\phi=2$.
The vortex lattice has honeycomb ordering at low $B_\phi$, but
as $B_\phi$ increases, the increasing strength of the vortex-vortex interactions
causes some of the vortices to shift out of the pinning sites.
(b) The corresponding anisotropy ratio $F^y_c/F^x_c$ vs $B_\phi$. 
(c) $F^x_c$ (open squares) and $F^y_c$ (filled circles)
vs $B_\phi$ for the same system at $B/B_\phi=3$.
Here the vortex lattice is triangular, as seen in Fig.~\ref{fig:5}(d), 
so as $B_{\phi}$ increases the
vortex-vortex interactions become stronger and increase the value of 
the interstitial vortex depinning threshold. 
(d) The corresponding anisotropy ratio $F^y_c/F^x_c$ vs $B_\phi$.
}
\label{fig:13}
\end{figure}

\begin{figure}
\includegraphics[width=3.5in]{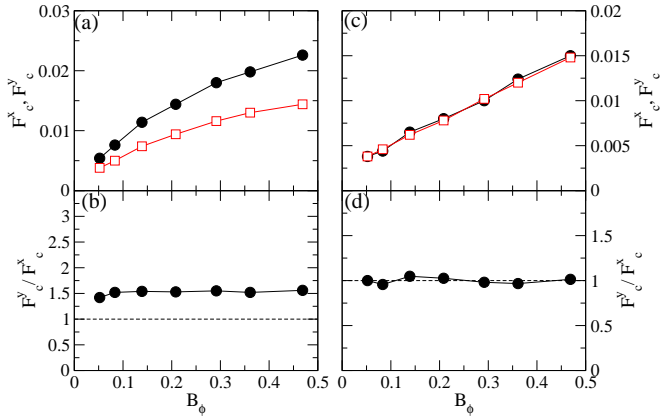}
\caption{(a) The depinning thresholds $F^{x}_{c}$ (open squares) and  
$F^{y}_{c}$ (filled circles) vs $B_\phi$
for a triangular pinning lattice system with $F_p=0.85$ and $R_p=0.35\lambda$
at $B/B_\phi=4$.
(b) The corresponding $F^y_c/F^x_c$ vs $B_\phi$.
(c) $F^x_c$ (open squares) and $F^y_c$ (filled circles) vs $B_\phi$ for the same
system at $B/B_\phi=5$.  
(d) The corresponding $F^y_c/F^x_c$ vs $B_\phi$ showing that the depinning is
isotropic.
}
\label{fig:14}
\end{figure}

\subsection{Effect of Changing $B_{\phi}$}

We next examine the depinning thresholds and anisotropy at different  
matching fields for the triangular pinning lattice with
fixed $F_{p} = 0.85$ but with increasing $B_\phi$, achieved
by increasing the density of pinning sites $n_p$.
The vortex-vortex interactions become more important 
for higher values of $B_{\phi}$.    
For $B/B_{\phi} = 2$, illustrated in Fig.~\ref{fig:13}(a,b), 
both $F^x_c$ and $F^y_c$ show a peak feature,
while the anisotropy remains near $F^y_c/F^x_c=2.0$.
The depinning thresholds increase with increasing $B_{\phi}$ 
for low $B_\phi$ due to the fact that the 
increasing strength of the vortex-vortex
interactions raises the interstitial pinning barriers.
The threshold does not continue to monotonically increase with increasing
$B_\phi$ since the vortex configuration at $B/B_\phi=2$ is a honeycomb
lattice.  As a result, the elastic energy cost of maintaining the honeycomb
structure increases with increasing $B_\phi$, and for 
$B_\phi>0.2\phi_0/\lambda^2$, a
portion of the vortices shift out of the pinning sites to form a distorted
triangular lattice similar to that seen for low $F_p$ in Fig.~\ref{fig:11}(c).
This causes a drop in the depinning thresholds with increasing $B_\phi$.
We expect similar behavior for other matching fields 
at which an ordered but non-triangular vortex lattice forms. 

In Fig.~\ref{fig:13}(c) we plot $F^x_c$ and $F^y_c$ versus $B_{\phi}$ 
for the same system with $B/B_\phi= 3$, and in Fig.~\ref{fig:13}(d) 
we show the corresponding $F^y_c/F^x_c$.
At this matching field, the vortex lattice is triangular 
as seen in Fig.~\ref{fig:5}(d).  Therefore, the vortex positions do not
shift as $B_{\phi}$ increases,
unlike the case for $B/B_\phi=2$,
and the depinning thresholds increase monotonically with increasing
$B_\phi$.
The anisotropy does not vary strongly with $B_\phi$.
In Fig.~\ref{fig:14}(a,b) 
we show $F^x_c$, $F^y_c$, and $F^y_c/F^x_c$ versus $B_\phi$ for the
same system at
$B/B_{\phi} = 4$, where 
the vortex lattice is triangular as indicated in Fig.~\ref{fig:6}(a). 
The depinning thresholds increase monotonically 
with increasing $B_{\phi}$ while the anisotropy remains constant at 
$F^y_c/F^x_c \approx 1.54$.
At $B/B_{\phi} = 5$, 
where Fig.~\ref{fig:6}(d) shows that the vortex lattice is disordered,
the depinning thresholds are isotropic and increase monotonically
with increasing $B_\phi$,
as illustrated in Fig.~\ref{fig:14}(c,d).
For any matching field $B/B_\phi$,
if $B_{\phi}$ is increased above the range of values considered here, 
multiple vortex pinning by the pinning sites eventually occurs
when the vortex lattice constant $a$ becomes of the order of 
the pinning radius $R_{p}$ 
and very little distortion of the vortex lattice is required to shift the 
vortices into the pinning sites. The occurrence of 
multiple vortex pinning would 
alter both the depinning thresholds and the anisotropy.         

\begin{figure}
\includegraphics[width=3.5in]{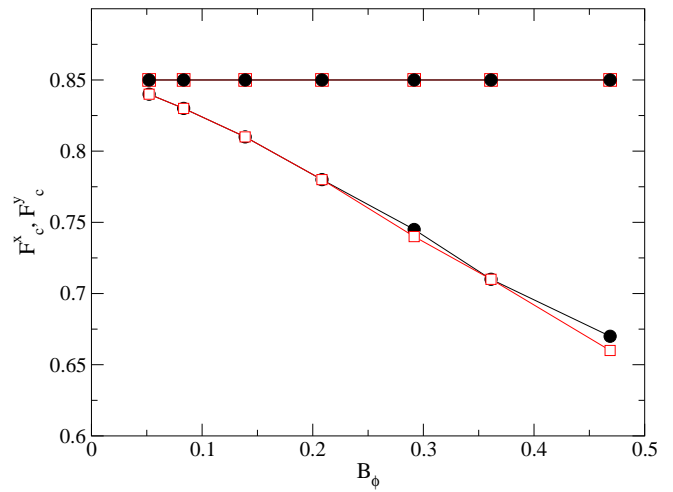}
\caption{
The depinning thresholds $F^{x}_{c}$ (open squares) and $F^{y}_{c}$ 
(filled circles) vs $B_{\phi}$ for the
system in Fig.~\ref{fig:2} with $F_p=0.85$. 
Lower curves: $B/B_{\phi} = 0.5$, showing that the depinning is isotropic and
monotonically decreasing with increasing $F_p$. 
Upper curves: $B/B_{\phi} = 1.0$, where $F_{c} = F_{p}$.
}
\label{fig:15}
\end{figure}

\subsection{Anisotropy for $B/B_\phi \leq 1$}

For $B/B_\phi > 1$, interstitial vortices are present
and the depinning threshold is determined by the 
strength of the vortex-vortex interactions,
provided that the pinning is strong enough to produce a plastic depinning
transition.
For 
$B/B_\phi \leq 1$, 
the depinning threshold is controlled by a combination of 
the vortex-vortex interaction strength and the strength of the 
pinning sites. 
For a triangular pinning array at the matching and submatching fields of 
$B/B_{\phi} = 1$, 1/3, and $1/4$, the
vortex lattice is triangular and every vortex is trapped by a pinning
site.
In this case, the depinning thresholds are
determined by the maximum force exerted by the pinning sites 
and are independent of the direction
of the external drive, so the depinning is isotropic. 
This is in agreement with previous work, where an isotropic depinning
threshold was observed for vortices in triangular pinning arrays
at $B/B_\phi=1$ \cite{Wu,Cao}.
In Fig.~\ref{fig:15} we plot $F^x_c$ and $F^y_c$ versus $B_\phi$ 
at $B/B_{\phi} = 0.5$ and $1.0$ 
for a system with $F_{p} = 0.85$. 
The depinning thresholds are isotropic for both fields, and
at $B/B_\phi=1$, 
$F_{c} = F_{p}$ as expected. 
At $B/B_{\phi} = 0.5$, the vortex lattice is disordered
since a triangular vortex lattice cannot match the pinning array
at this filling, as shown in Fig.~11(c) of Ref.~\cite{Reichhardt2001}. 
Although all of the vortices are pinned, due to the disorder of the 
vortex structure, some vortices experience stronger vortex-vortex repulsion
from neighboring vortices than other vortices, and as a result the
vortex-vortex interactions do not cancel 
at $B/B_\phi=0.5$ as they do at $B/B_\phi=1$.
This lowers the depinning thresholds and causes both $F^x_c$ and $F^y_c$ to
decrease with increasing $B_\phi$ due to the dependence of the depinning
force on the vortex-vortex interaction strength.
In general, we observe little or no anisotropy 
in the depinning thresholds for $B/B_\phi < 1$. 
If multiple vortex pinning occurs at  $B/B_\phi > 1$, so that no 
interstitial vortices are present,
we expect that the
same type of depinning phenomena seen for $B/B_\phi \leq 1$ 
will appear 
and there will be little anisotropy. 
This suggests that the appearance of anisotropic depinning forces and 
the existence
of voltage-current curves with different values of $dV/dF_d$ in different
directions 
are
indicators of the presence of interstitial vortices in the system. 

\begin{figure}
\includegraphics[width=3.5in]{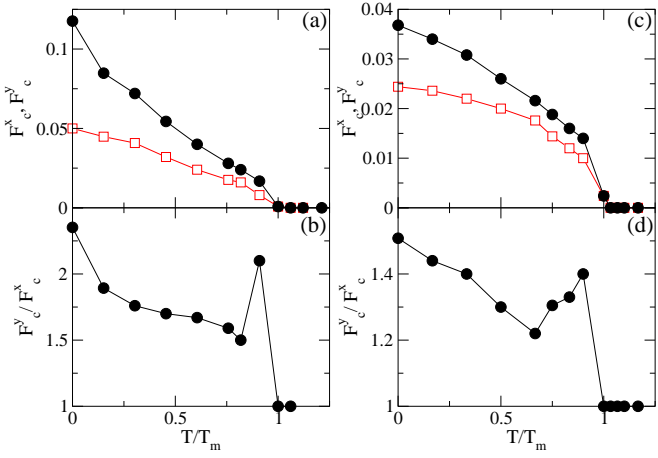}
\caption{(a) The depinning thresholds 
$F^{x}_{c}$ (open squares) and
$F^{y}_{c}$ (filled circles) 
vs $T/T_{m}$ for  a triangular pinning lattice system
with $F_p=1.25$ and $B/B_{\phi} = 2$.
Here $T_{m}$ is the
temperature associated with the onset of vortex diffusion at 
$F_{D}= 0$. 
(b) The corresponding anisotropy ratio $F^{y}_{c}/F^{x}_{c}$
vs $T/T_m$. 
(c) $F^x_c$ (open squares) and $F^y_c$ (filled circles) vs $T/T_m$ for
the same system at $B/B_{\phi} = 4$. 
(d) The corresponding $F^y_c/F^x_c$ vs $T/T_m$.
}
\label{fig:16}
\end{figure}

\subsection{Temperature Effects}

We next consider how robust the anisotropy is to thermal fluctuations. 
We anneal the system down to a finite temperature and then 
measure the depinning forces in the $x$ and $y$-directions. 
The temperature is given in terms of the melting temperature $T_{m}$
at which the vortices begin to diffuse significantly and the system is
in a molten state.
In Fig.~\ref{fig:16}(a) we plot $F^{x}_{c}$ and $F^{y}_{c}$ 
versus $T/T_{m}$ for a triangular pinning lattice system with $F_p=1.25$ at
$B/B_{\phi} = 2$, and in Fig.~\ref{fig:16}(b) we show the corresponding 
anisotropy $F^y_c/F^x_c$.
Both depinning thresholds monotonically decrease with increasing temperature 
and reach zero at $T/T_{m} = 1.0$.
The anisotropy decreases with increasing temperature for low temperatures;
however, just below the melting temperature, $F^y_c/F^x_c$ passes through
a peak.
A similar trend is seen for $B/B_{\phi} = 4$, as shown in
Fig.~\ref{fig:16}(c,d). 
These results indicate that the anisotropy should be robust 
against temperature. This agrees with the experiments in Ref.~\cite{Cao},
which
found that the anisotropy became more pronounced at higher temperatures. 
The enhancement of the anisotropy at higher temperatures could also result
from the presence of weak random intrinsic pinning in the sample, which would
reduce the magnitude of the anisotropy but which can 
be washed out by thermal
fluctuations.

\begin{figure}
\includegraphics[width=3.5in]{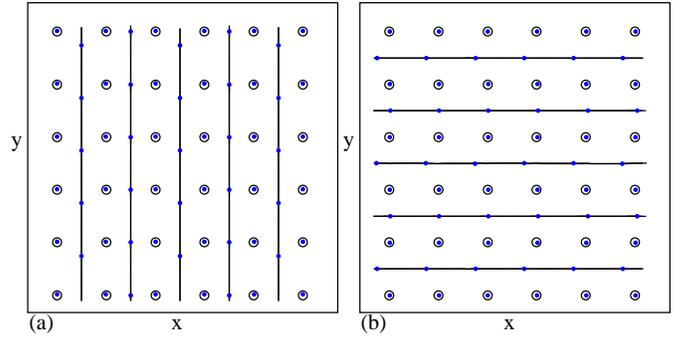}
\caption{
Vortex positions (black dots), pinning site locations (open circles), 
and vortex trajectories (black lines) 
just above depinning for a system with a square pinning array 
at $F_{p} = 0.86$, $R_{p} = 0.35\lambda$, and $B_{\phi} = 0.0625\phi_0/\lambda^2$
at $B/B_\phi=2$.
(a) $y$-direction driving. (b) $x$-direction driving.
The same type of vortex motion occurs for both directions of drive.    
}
\label{fig:17}
\end{figure}

\begin{figure}
\includegraphics[width=3.5in]{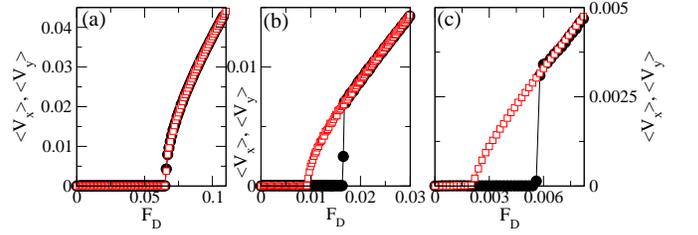}
\caption{Velocity-force curves 
$\langle V_x\rangle$ vs $F_D$ for $x$-direction depinning (open squares) 
and $\langle V_y\rangle$ vs $F_D$ for $y$-direction depinning (filled circles) 
in the square pinning lattice system from Fig.~\ref{fig:17}.
(a) $B/B_{\phi} = 2$, where the depinning is isotropic. 
(b) $B/B_{\phi} = 4$, with anisotropic
depinning. 
(c) $B/B_{\phi} = 12$, with anisotropic depinning.     
}
\label{fig:18}
\end{figure}

\begin{figure}
\includegraphics[width=3.5in]{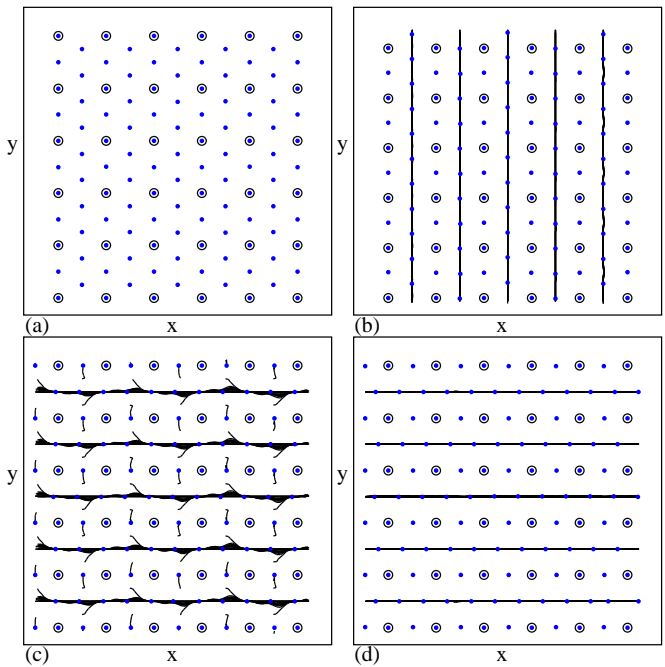}
\caption{
The vortex positions (black dots), pinning site locations (open circles), 
and vortex trajectories (black lines) for the 
square pinning lattice system in Fig.~\ref{fig:18} at $B/B_{\phi} = 4$. 
(a) The $F_{D} = 0$ state where a triangular vortex lattice forms
that is aligned in the $y$-direction. 
(b) The vortex trajectories just above depinning for driving in the 
$y$-direction, showing 
one-dimensional channels of vortices moving between adjacent columns
of pinning sites. 
(c) The vortex trajectories just above depinning 
for driving in the $x$-direction. 
Here, every other column of vortices shifts in the $y$-direction so that
every other vortex can join
one-dimensional rows of vortices flowing between
adjacent pinning site rows, while the remaining interstitial vortices are
trapped between neighboring pinning sites.
(d) The same as panel (c), but at a later time when the system has reached
steady-state flow.
 }
\label{fig:19}
\end{figure}

\section{Anisotropy in Square Pinning Arrays} 

For square pinning arrays, we find that the depinning thresholds at most
fields are isotropic. This is a result of the fact that the same type of
vortex motion occurs in both directions for most fields, as illustrated
for $B/B_\phi=2$ in Fig.~\ref{fig:17}.
In Fig.~\ref{fig:18}(a) we show the isotropic velocity-force curves 
at $B/B_{\phi} = 2$ for $x$ and $y$-direction driving.
Since the perpendicular directions of the square pinning array are 
identical, unlike the perpendicular directions of the triangular pinning
array, 
it is not surprising that most matching fields have the same depinning 
thresholds in both driving directions for the square array.  
Nevertheless, strongly anisotropic depinning occurs at $B/B_{\phi} = 4$ 
and $12$, as shown in Fig.~\ref{fig:18}(b,c). 

In Fig.~\ref{fig:19}(a) we plot the vortex positions and pinning site
locations for $B/B_{\phi} = 4$, where anisotropic depinning occurs. 
Here a triangular vortex lattice that is aligned with the $y$-axis forms.
Under $y$-direction driving, the interstitial vortices flow in one-dimensional
channels between adjacent columns of pinning sites, as seen in 
Fig.~\ref{fig:19}(b) where $2/3$ of the interstitial vortices have
depinned.
A simple channeling motion cannot occur for $x$ direction driving, so
$F^x_c$ is much higher than $F^y_c$.
Just above the depinning transition for $x$-direction driving,
as illustrated in Fig.~\ref{fig:19}(c),
every other column of vortices shifts in the $y$ direction in order to permit
every other vortex to join a one-dimensional flowing channel passing
between adjacent rows of pinning sites.  The remaining interstitial
vortices remain trapped between the pinning sites.
After this rearrangement, $2/3$ of the interstitial vortices flow in
the steady-state one-dimensional channels shown in Fig.~\ref{fig:19}(d), which
are similar to the channels that form for $y$-direction driving.
The ground state shown in Fig.~\ref{fig:19}(a) is degenerate, and a state
with the vortex lattice aligned along the $x$-direction has equal energy.
Thus, the realignment process seen in Fig.~\ref{fig:19}(c) is simply a
shift of the vortices into the other ground state
prior to the onset of flow.
Figure~\ref{fig:18}(b) shows that $F^x_c$ is approximately 1.75 times higher
than $F^y_c$.
Due to the symmetry of the square lattice, either the $x$ or the $y$-direction
can show a higher depinning threshold depending on the initial configuration
of the vortex lattice.  This is distinct from the triangular pinning lattice,
where the higher depinning force always occurs in the same direction at a
given field.
If the annealing process for the square pinning lattice 
is repeated with different initial 
conditions, the vortex lattice has a 50\% chance of aligning with the 
$x$-direction, in which case the anisotropy will be reversed from that
shown in Figs.~\ref{fig:18}, \ref{fig:19}. 
If the driving force is cycled,
the velocity-force curve in the hard driving direction is hysteretic during
the first cycle due to the realignment effect, 
while there is no hysteresis for the easy driving direction or for subsequent
cycles in the initially hard driving direction.

\begin{figure}
\includegraphics[width=3.5in]{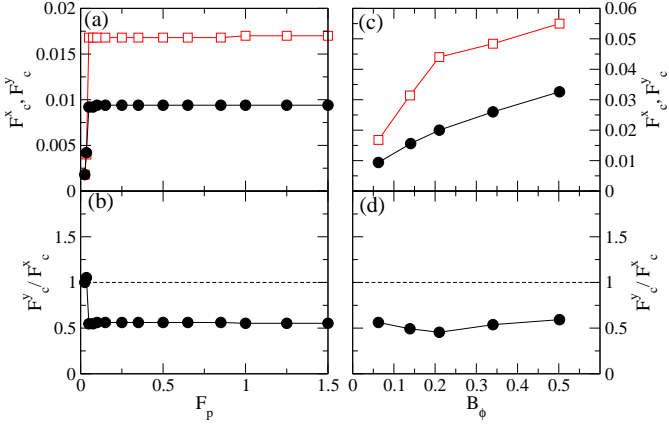}
\caption{
(a) $F^{x}_{c}$ (open squares) and $F^{y}_{c}$ (filled circles) vs $F_p$
for the system in Fig.~\ref{fig:18} at $B/B_{\phi} = 4$ with 
the starting configuration shown in Fig.~\ref{fig:19} that produces a
higher depinning threshold in the $x$-direction. 
(b) The corresponding anisotropy ratio $F^{y}_{c}/F^{x}_{c}$
vs $F_p$. 
For small $F_{p}$ the system depins elastically. 
(c) $F^x_c$ (open squares) and $F^y_c$ (filled circles) vs $B_\phi$ for
the same system at $B/B_\phi=4$ and $F_p=0.85$.
(d) The corresponding $F^y_c/F^x_c$ vs $F_p$.
}
\label{fig:21}
\end{figure}

In Fig.~\ref{fig:21} we illustrate the effect of changing $F_p$ and $B_\phi$ on the
depinning anisotropy for the square pinning lattice sample with $B/B_\phi=4$.
We show a case where the initial vortex lattice orientation is along
the $y$-axis, as in Fig.~\ref{fig:19}(a), which gives $F^x_c>F^y_c$.
Figure~\ref{fig:21}(b) indicates that the anisotropy saturates
at $F^y_c/F^x_c \approx 0.57$ for $F_{p} >0.075$, while
Fig.~\ref{fig:21}(a) shows that the depinning thresholds
also saturate above this value of $F_p$.
For $F_{p} < 0.075$, the depinning is elastic and the 
anisotropy is lost. 
For fixed $F_p=0.85$ 
the depinning thresholds increase monotonically with 
increasing $B_{\phi}$, 
as seen in Fig.~\ref{fig:21}(c),
while in Fig.~\ref{fig:21}(d) the anisotropy 
passes through a shallow extremum of $F^y_c/F^x_c=0.5$.

\begin{figure}
\includegraphics[width=3.5in]{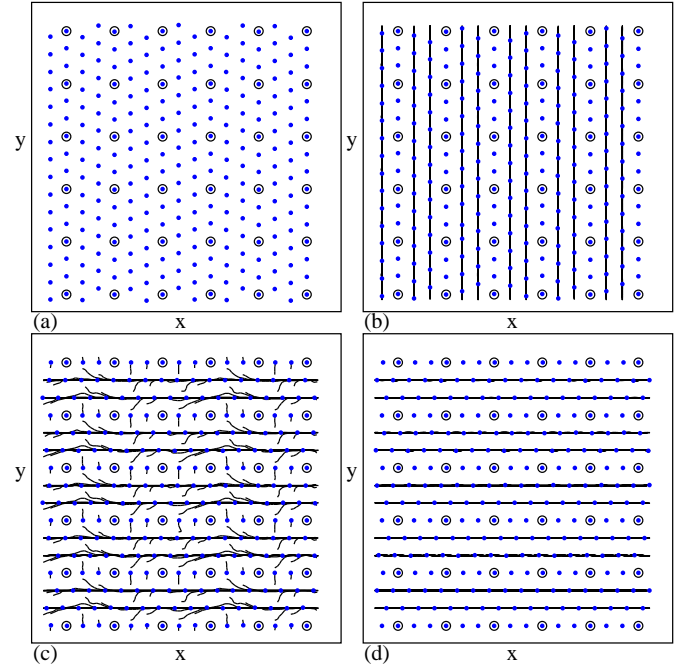}
\caption{
The vortex positions (black dots), pinning site locations (open circles), 
and vortex trajectories (black lines) for the system in 
Fig.~\ref{fig:18} at $B/B_{\phi} = 12.$
(a) The partially ordered vortex configuration at $F_{D} = 0$. 
The vortex structure has a smectic-type ordering in which all of the
topological defects are aligned in the same direction. 
(b) The vortex trajectories just above depinning for driving in the 
$y$-direction showing two columns of interstitial vortices moving 
between each pair of pinning columns and two interstitial vortices 
trapped between adjacent pinning sites.
(c) The vortex trajectories just above depinning for driving in the 
$x$-direction. A portion of the
interstitial vortices shift in the $y$-direction in order to create ordered
rows of flowing vortices and pairs of trapped interstitial vortices
between adjacent pinning sites. 
(d) The steady-state channels that form for $y$-direction driving.
The general structural rearrangement of the interstitial
vortices for depinning in the $x$-direction 
is similar to that seen at $B/B_{\phi} = 4$ in Fig.~\ref{fig:19}(c,d).
}
\label{fig:20}
\end{figure}

We find a similar anisotropic depinning behavior for $B/B_\phi=12$.
The vortex configurations for this field at $F_{D} = 0$ 
are illustrated in Fig.~\ref{fig:20}(a). 
The vortex lattice is aligned in the $y$ direction, but just as at $B/B_\phi=4$,
there are two degenerate ground states, and a vortex lattice that is aligned
in the $x$-direction would have the same energy.
There are two columns of interstitial vortices between adjacent
columns of pinning sites.
The overall vortex lattice structure is not triangular and there are
dislocations present in the lattice.
Along certain columns, 
neighboring interstitial vortices in neighboring columns lie along a line
tilted by $-23^\circ$ with respect to the $x$-axis, while along other columns,
neighboring interstitial vortices lie along a line tilted by $+23^\circ$ with
respect to the $x$-axis.
The dislocations in the vortex lattice 
are all aligned in the same direction, 
resulting in
a smectic structure.
This smectic state for the square pinning array has not been observed 
in previous work. 
There are two possible low-energy orientations for the vortex lattice,
just as in the $B/B_\phi=4$ case: the $y$-axis orientation shown in 
Fig.~\ref{fig:20}(a), or the same state rotated by $90^\circ$ and aligned
with the $x$-axis.
For depinning in the $y$-direction, Fig.~\ref{fig:20}(b) shows that 
the two columns of interstitial vortices 
depin into flowing one-dimensional channels, while two interstitial vortices 
remain trapped between adjacent pairs of pinning sites. 
For driving along the $x$-direction, the same type of lattice reorientation
found for $B/B_\phi=4$ occurs at $B/B_\phi=12$, as illustrated in
Fig.~\ref{fig:20}(c,d).
The columns of interstitial vortices shift in such a way that two 
one-dimensional rows of interstitial vortices form, while two interstitial
vortices remain trapped between adjacent pairs of pinning sites.
The final vortex configuration for $x$-direction driving, in 
Fig.~\ref{fig:20}(d), is a rotated version of the configuration for 
$y$-direction driving seen in Fig.~\ref{fig:20}(b).
The same hysteretic voltage-current response should occur for driving in the
hard direction at $B/B_\phi=12$ as at $B/B_\phi=4$.
In Fig.~\ref{fig:18}(c) the velocity-force curves 
at $B/B_\phi=12$ show an anisotropy 
$F^y_c/F^x_c \approx 0.4.$ 
The more pronounced anisotropy at $B/B_\phi=12$ compared to 
$B/B_\phi=4$ is due to the fact that the overall structure 
at $B/B_\phi=12$ is smectic and
the dislocations are aligned in the $y$-direction, further decreasing
the depinning force along this direction.  

We expect that the depinning anisotropy in the square pinning arrays will be
more difficult to observe experimentally than the anisotropy in the triangular
pinning arrays due to the existence of 
twofold-degenerate ground states for the square system.
If other forms of quenched disorder, such as intrinsic pinning, 
are present and the system is large, domains of 
the two different orientations could form which would render 
the depinning thresholds isotropic. On the other hand, we found that
an applied drive can readily align the vortex lattice structures in the
driving direction
at $B/B_{\phi} = 4$ and $12$. 
Therefore, it may be possible to prepare the system in an aligned state using
an external drive, and then measure the anisotropy of the depinning forces
starting from this aligned state.  This 
procedure should permit the anisotropy in the square pinning lattice system to
be observed experimentally.

\begin{figure}
\includegraphics[width=3.5in]{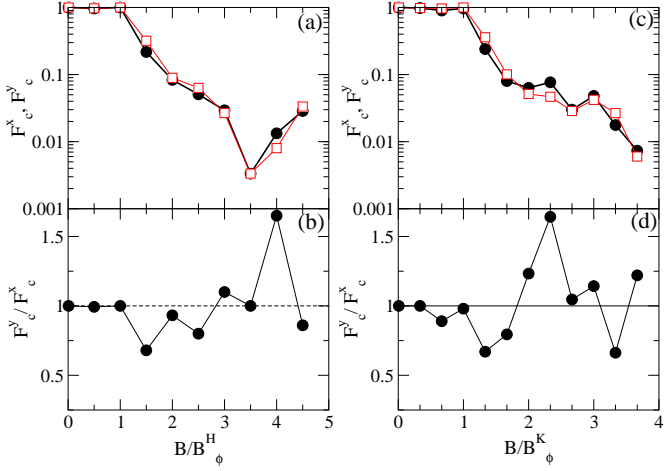}
\caption{
(a) $F^x_c$ (open squares) and $F^y_c$ (filled circles) for a honeycomb pinning
lattice as a function of $B/B^H_\phi$, the honeycomb matching field, in
a sample with $F_p=0.85$, $R_p=0.35\lambda$, and $n_p=0.194/\lambda^2$.
(b) The corresponding $F^{y}_{c}/F^{x}_{c}$ 
vs $B/B^H_\phi$ shows that the 
anisotropy exhibits several reversals as a function of field.
(c) $F^x_c$ (open squares) and $F^y_c$ (filled circles) for a kagom{\' e}
pinning lattice as a function of $B/B^K_\phi$, the kagom{\' e} matching
field.
(d) The corresponding $F^y_c/F^x_c$ vs $B/B^K_\phi$ shows 
that several reversals of the anisotropy also occur for
the kagom{\' e} pinning array.
}
\label{fig:22}
\end{figure}

\begin{figure}
\includegraphics[width=3.5in]{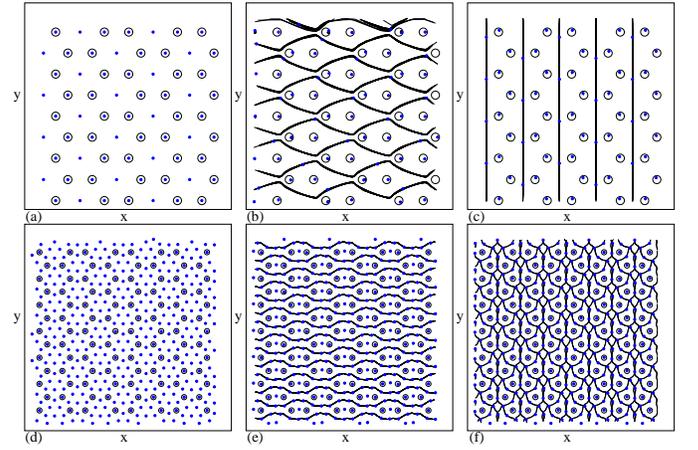}
\caption{
Vortex positions (black dots), pinning site locations (open circles), 
and vortex trajectories (black lines)
for the honeycomb pinning array in Fig.~\ref{fig:22}.
(a) At $B/B^{H}_{\phi} = 1.5$ and $F_D=0$ 
there is one vortex per large interstitial site.
(b) Vortex  trajectories just above depinning 
for driving in the $y$-direction 
at $B/B^H_\phi=1.5$
showing one-dimensional paths between the pinning sites. At
this field, $F^y_c/F^x_c<1$.
(c) Vortex trajectories just above depinning for driving
in the $x$-direction at $B/B^H_\phi=1.5$. 
The interstitial vortices flow in winding paths 
around the occupied pinning sites. 
(d) The vortex configurations at $B/B^{H}_{\phi} = 4$ for $F_{D} = 0$. 
Here a portion of the interstitial vortices align in the $x$-direction.
(e) Vortex trajectories just above depinning for driving in the $x$-direction
at $B/B^H_\phi=4$.
In addition to moving channels of interstitial vortices, there are
some interstitial vortices that remain pinned between adjacent occupied
pinning sites.
(f) Vortex trajectories just above depinning for driving in the 
$y$-direction at $B/B^H_\phi=4$ show a complex periodic pattern with all
of the interstitial vortices moving.       
}
\label{fig:24}
\end{figure}

\section{Anisotropy in Honeycomb and Kagom{\' e} Pinning Arrays}    

We next examine anisotropy in honeycomb and 
kagom{\' e} pinning arrays. 
Since the honeycomb and kagom{\' e} arrays can be constructed
by removing selected pinning sites from a triangular array, 
it might be expected that the anisotropy would follow the
same trend found for the triangular pinning  arrays. 
In particular, one could expect that the depinning threshold 
would always be higher in the $y$-direction.
Instead, we find that the honeycomb and kagom{\' e} pinning 
arrangements show an anisotropy that undergoes reversals as a function of the
applied magnetic field. 
In Fig.~\ref{fig:22}(a) we plot $F^x_c$ and $F^y_c$ as a function
of $B/B^H_\phi$ for a honeycomb pinning 
system with $F_p=0.85$, $R_p=0.35\lambda$,
and $n_p=0.194/\lambda^2$.  Here, $B^H_\phi$ is the matching field for
a honeycomb lattice constructed from a triangular lattice
with matching field $B_\phi$, and we have $B^H_\phi=2/3B_\phi$ \cite{Molecular}.
Figure \ref{fig:22}(b) shows 
$F^{y}_{c}/F^{x}_{c}$ versus $B^{H}_{\phi}$ for the honeycomb pinning array. 
For $1 < B/B^H_\phi < 3$, $F^y_c/F^x_c<1$ and the depinning threshold is higher
for $x$-direction driving.
This anisotropy is reversed from that seen in the triangular pinning arrays. 
The reversal can be understood by examining the vortex positions 
at $B^{H}_{\phi} = 1.5$ in Fig.~\ref{fig:24}(a). 
Each interstitial vortex is located
at the position where the pinning site was removed
from the triangular lattice in order to create the honeycomb
lattice. 
There is a pin-free channel of motion which the interstitial vortices can
follow for driving in the $y$-direction, as seen in Fig.~\ref{fig:24}(c).
For $x$-direction driving, the path of the interstitial vortex is blocked by
pinned vortices, creating a much stronger barrier for depinning, and once
the vortices begin to move, they follow the winding paths illustrated
in Fig.~\ref{fig:24}(b).

\begin{figure}
\includegraphics[width=3.5in]{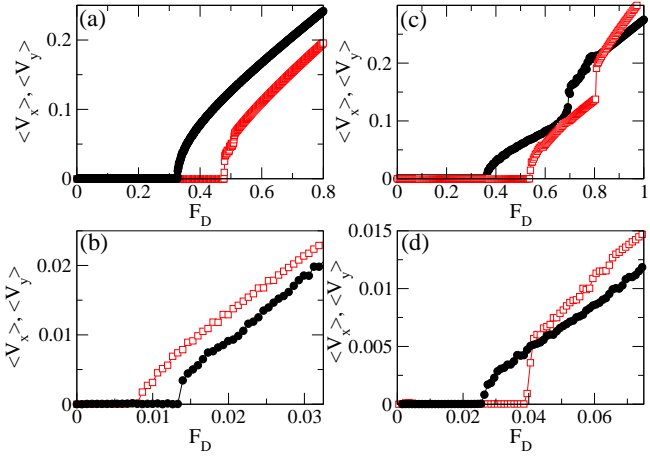}
\caption{
The velocity-force curves from the system in 
Fig.~\ref{fig:22}. Open squares: $\langle V_x\rangle$
vs $F_D$; filled circles:
$\langle V_y\rangle$ vs $F_D$.
(a) The honeycomb pinning array at $B/B^{H}_{\phi} = 1.5$ where 
$F^y_c/F^x_c<1$.
(b) The honeycomb pinning array at $B/B^{H}_{\phi} = 4$ showing 
that the anisotropy has reversed and $F^y_c/F^x_c>1$.
(c) The kagom{\' e} pinning array at $B/B^{K}_{\phi} = 4/3$. 
(d) The kagom{\' e} pinning array at $B/B^{K}_{\phi} = 10/3$. 
Here $dV_y/dF_D<dV_x/dF_D$, so there is a crossing of the velocity-force
curves.
}
\label{fig:23}
\end{figure}

In Fig.~\ref{fig:23} we plot $\langle V_x\rangle$ and $\langle V_y\rangle$
versus $F_D$ for driving in the $x$ and $y$-directions, respectively.
The depinning in the honeycomb pinning array 
is anisotropic at both $B/B^{H}_{\phi} = 1.5$, shown in Fig.~\ref{fig:23}(a),
and at $B/B^{H}_{\phi} = 4$, shown in Fig.~\ref{fig:23}(b).  
For $2.5 \leq B/B^{H}_{\phi} \leq 4$  the
anisotropy is reversed compared to the lower fields.
This is due to the formation of $n$-mer states 
in the large interstitial spaces of the honeycomb lattice
which permit a portion of the interstitial vortices to be aligned
in the $x$-direction between the pinning rows. 
In Fig.~\ref{fig:24}(d) we illustrate 
the $F_{D} = 0$ vortex configuration at $B/B^{H}_{\phi} = 4$, 
where the interstitial vortices form triangular shapes 
within the large interstitial sites. 
The vortex at the top of each 
interstitial triangle forms a nearly one-dimensional 
channel in the $x$-direction with the vortices at the base of the adjacent
interstitial triangles, permitting easy depinning in the $x$-direction into
the flow pattern shown in Fig.~\ref{fig:24}(e).
For depinning in the $y$-direction, where the threshold is higher, the complex
but ordered flow pattern in Fig.~\ref{fig:24}(f) appears.       
For $B^{H}_{\phi} = 4.5$, the anisotropy reverses again, as seen
in Fig.~\ref{fig:22}. 

Figure~\ref{fig:22}(c,d) shows that similar anisotropy reversals occur for 
depinning in kagom{\' e} pinning arrays, where
$B^K_\phi$ is the 
kagom{\' e} matching field for a pinning lattice
constructed from a triangular lattice with matching field $B_\phi$ and
$B^K_\phi=3/4B_\phi$.
For 
$B/B^{K}_{\phi} \leq 5/3$, Fig.~\ref{fig:22}(d) indicates that $F^y_c/F^x_c<1$,
but that this pattern reverses several times for increasing $B/B^K_\phi$.
Just as in the honeycomb pinning array, in the kagom{\' e} pinning array
the reversals originate from the formation and alignment of interstitial vortex
$n$-mer states in the large interstitial sites.
In general, the anisotropy is smaller for the honeycomb and 
kagom{\' e} pinning arrays than for the triangular or square pinning arrays.   

\begin{figure}
\includegraphics[width=3.5in]{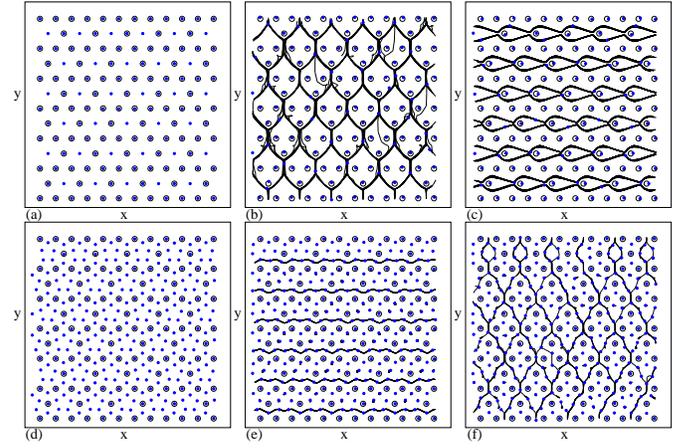}
\caption{Vortex positions (black dots), 
pinning site locations (open circles), and 
vortex trajectories (black lines)
for the kagom{\' e} pinning array. 
(a) The $F_D=0$ state at $B/B^{K}_{\phi} = 4/3$, 
where there is one vortex per large interstitial site.   
(b) The vortex trajectories just above depinning 
for driving in the $y$-direction at $B/B^K_\phi=4/3$ shows the formation of
an asymmetric pattern that encircles groups of three pinning sites. 
(c) The vortex trajectories just above depinning 
for driving in the $x$-direction at $B/B^K_\phi=4/3$
shows a winding channel of interstitial vortices between the pinning rows.
(d) The $F_{D} = 0$ state at $B/B^{K}_{\phi} = 10/3$.
(f) The vortex trajectories just above depinning for driving
in the $x$-direction at $B/B^K_\phi=10/3$
shows that a portion of the interstitial vortices 
move continuously through the system.   
(e) The vortex trajectories just above depinning for driving
in the $y$-direction at $B/B^K_\phi=10/3$ 
shows a winding pattern of interstitial vortices. 
Here the vortex motion occurs in intermittent pulses 
rather than as a continuous flow.
}
\label{fig:25}
\end{figure}

We illustrate the $F_D=0$ state for $B/B^K_\phi=4/3$ in 
Fig.~\ref{fig:25}(a), 
where there is one vortex per large interstitial site.  
In Fig.~\ref{fig:25}(b) 
we show the vortex trajectories just above depinning 
for driving in the $y$-direction at $B/B^{K}_{\phi} = 4/3$. 
Each interstitial vortex diverts to the right or left around a
pinned vortex, forming
large asymmetric patterns of vortex flow around trios of occupied pins.
The flow is slightly disordered, since in some cases two moving 
interstitial vortices approach the same pinned vortex simultaneously, 
causing one of the vortices to move outside of the flow pattern temporarily.
For $x$-direction driving 
at $B/B^K_\phi=4/3$, shown in Fig.~\ref{fig:25}(c),
the interstitial vortices flow around individual pinning sites in the sparse
pinning rows, creating an asymmetric sinusoidal channel pattern.
Figure~\ref{fig:23}(c) indicates that at this field, $F^y_c/F^x_c<1$ and
depinning is easiest along the $y$-direction.
A similar anisotropy occurs for $B/B^{k}_{\phi} = 10/3$, as shown in
Fig.~\ref{fig:23}(d).
In Fig.~\ref{fig:25}(d) 
we illustrate the $F_{D} = 0$ state at $B/B^K_\phi=10/3$, 
where the large interstitial sites capture five vortices 
which form a pentagon structure. 
For depinning in the $x$-direction at this field, 
only a portion of the interstitial 
vortices move in winding channels between the
pinning rows while the large interstitial sites all capture three vortices,
as seen in Fig.~\ref{fig:25}(e). 
The interstitial vortices which are not part of the channeling flow
still undergo a small circular motion as the 
flowing interstitial vortices move past. 
At higher drives, all of the interstitial vortices depin
and a step appears in the velocity-force curve.
For depinning in the $y$-direction at $B/B^K_\phi=10/3$, 
shown in Fig.~\ref{fig:25}(f), the trajectories of the moving
interstitial vortices are much more tortuous.
Although a larger number of interstitial vortices spend at least part of the
time flowing for depinning in the $y$-direction than for depinning in the
$x$-direction, Fig.~\ref{fig:23}(d) shows that 
$dV_y/dF_D<dV_x/dF_D$.
This is because the vortex motion in 
Fig.~\ref{fig:25}(f) is not continuous.  Instead, the interstitial
vortices hop by one lattice constant and then repin so that a pulse of motion
passes though the system. 
For depinning in the $x$-direction, shown in Fig.~\ref{fig:25}(e), 
the vortices are continuously moving through the system.            

\section{Summary}

We have shown that when interstitial vortices are present for fields 
beyond the first matching field in regular artificial pinning arrays, the
transport response is anisotropic.
For triangular pinning arrays, we find that the depinning thresholds are 
always higher in the $y$-direction, defined to be perpendicular to a symmetry 
axis of the pinning lattice. 
This is in general agreement with previous numerical and experimental studies
up to the third matching field for triangular pinning arrays. 
We show that the anisotropy also occurs for higher matching fields and that
at certain matching fields the depinning is isotropic. 
The degree of anisotropy can be controlled by changing the strength of the
pinning. 
For weak pinning, when the depinning transition is elastic and all of 
the vortices depin simultaneously, the anisotropy is reduced or 
destroyed. 
For strong pinning, the depinning is isotropic for matching fields 
at which the vortex lattice does not order. 
We find that the velocity-force curves can have 
different slopes for driving in the different directions 
depending on the number of interstitial vortices that initially depin.
In some cases, although the depinning threshold is lower in one direction, 
the slope of the velocity-force curve is also lower in that direction, producing
a crossing in the velocity-force curves for the two directions of
driving.
For fields at or below the first matching field, the anisotropy is strongly 
reduced, implying that the anisotropy should disappear if multiple vortex
pinning rather than interstitial vortex pinning 
occurs above the first matching field. 
The vortex dynamics can be distinctly different for the two 
driving directions, with ordered flow states occurring for driving
in one direction and disordered flow states appearing for driving 
in the other direction. 
The anisotropy is robust against temperature fluctuations
and can be enhanced near the vortex lattice melting transition.  
For square pinning arrays, the two perpendicular driving directions
are symmetric; nevertheless, at certain matching fields where triangular or
smectic vortex structures form, a strong depinning anisotropy can occur.
The easy-flow direction for square lattices is not fixed but depends on which
of the two degenerate ground states is formed by the initial vortex lattice.
A sufficiently large applied drive realigns the vortex lattice and sets 
the easy-flow
direction in the direction of the drive.  If the initial vortex lattice formed
with domains of the two degenerate ground states, the domains can be 
eliminated and a pure ground state formed simply by sweeping the driving force.
The anisotropic depinning behavior of this pure state can then be 
probed with small drives.
Honeycomb and kagom{\' e} pinning arrays have a smaller anisotropy 
than that seen for the triangular and square arrays.
The anisotropy for the honeycomb and kagom{\' e} arrays 
shows a series of reversals as a function of field 
due to the formation of 
vortex molecular crystals, which have orientational degrees of freedom
that can lock to different angles.  

\acknowledgments{
This work was carried out under the auspices of the National Nuclear
Security Administration of the U.S. Department of Energy at
Los Alamos National Laboratory under Contract No.~DE-AC52-06NA25396.}


\begin{thebibliography}{99}
\bibitem{Fiory}
A.T.~Fiory, A.F.~Hebard, and S.~Somekh, Appl.~Phys. Lett. {\bf 32}, 73 (1978).

\bibitem{Metlushko}
V.V.~Metlushko, M.~Baert, R.~Jonckheere, V.V. Moshchalkov, 
and Y. Bruynseraede, 
Solid State Commun. {\bf 91}, 331 (1994).  

\bibitem{Baert}
M.~Baert, V.V.~Metlushko, R.~Jonckheere, V.V. Moshchalkov, 
and Y. Bruynseraede, Phys.~Rev.~Lett.~{\bf 74}, 3269 (1995);
M.~Baert, V.V.~Metlushko, R.~Jonckheere, V.V.~Moshchalkov, 
and Y.~Bruynseraede, Europhys.~Lett.~{\bf 29}, 157 (1995).    

\bibitem{Pannetier}
A.~Bezryadin and B.~Pannetier, J.~Low Temp. Phys. {\bf 102}, 73 (1996).  
 
\bibitem{Rosseel}
E.~Rosseel, M. Van Bael, M.~Baert, R.~Jonckheere, 
V.V.~Moshchalkov, and Y.~Bruynseraede, Phys.~Rev.~B {\bf 53}, R2983 (1996);
L.~Van Look, E.~Rosseel, M.J.~Van Bael, K.~Temst, V.V.~Moshchalkov, 
and Y.~Bruynseraede, Phys.~Rev.~B {\bf 60}, 
R6998 (1999). 

\bibitem{Harada}
K.~Harada, O.~Kamimura, H.~Kasai, T.~Matsuda, A.~Tonomura, 
and V.V.~Moshchalkov, Science {\bf 274}, 1167 (1996). 

\bibitem{Welp}
V.~Metlushko, U.~Welp, G.W.~Crabtree, Z.~Zhang, S.R.J.~Brueck, 
B.~Watkins, L.E.~DeLong, B.~Ilic, K.~Chung, and
P.J.~Hesketh, Phys.~Rev.~B {\bf 59}, 603 (1999).   

\bibitem{Kwok}
U.~Welp, Z.L. Xiao, J.S. Jiang, V.K. Vlasko-Vlasov, S.D. Bader,
G.W. Crabtree, J. Liang, H. Chik, and J.M. Xu, 
Phys.~Rev.~B {\bf 66}, 212507 (2002).

\bibitem{Martin}
J.I.~Mart{\' \i}n, M.~V{\' e}lez, J.~Nogu{\' e}s, 
and I.K.~Schuller, Phys.~Rev.~Lett.~{\bf 79} , 1929 (1997);
A. Hoffmann, L.~Fumagalli, N.~Jahedi, J.C.~Sautner, J.E.~Pearson, 
G.~Mihajlovic, and V.~Metlushko, Phys.~Rev.~B {\bf 77},
060506(R) (2008).    

\bibitem{Ketterson}
D.J.~Morgan and J.B.~Ketterson, Phys.~Rev.~Lett.~{\bf 80}, 3614 (1998). 

\bibitem{Zhukov}
A.A.~Zhukov, P.A.J.~de Groot, V.V.~Metlushko, 
and B. Ilic, Appl.~Phys.~Lett.~{\bf 83}, 4217 (2003). 

\bibitem{Raedts}
A.V.~Silhanek, S.~Raedts, M.J.~Van Bael, and V.V.~Moshchalkov, 
Phys.~Rev.~B {\bf 70}, 054515 (2004).  

\bibitem{Goran}
G.~Karapetrov, J.~Fedor, M.~Iavarone, D.~Rosenmann, and W.K.~Kwok, 
Phys.~Rev.~Lett.~{\bf 95}, 167002 (2005).  

\bibitem{Fasano}
Y.~Fasano and M.~Menghini, Supercond. Sci. Technol. {\bf 31}, 023001 (2008).  

\bibitem{Field}
S.B.~Field, S.S.~James, J.~Barentine, V.V.~Metlushko, G.~Crabtree, 
H.~Shtrikman, B.~Ilic, and S.R.J.~Brueck,
Phys.~Rev.~Lett.~{\bf 88}, 067003 (2002).   

\bibitem{Bending}
A.N.~Grigorenko, G.D. Howells, S.J. Bending, J. Bekaert, M.J. Van Bael,
L. Van Look, V.V. Moshchalkov, Y. Bruynseraede, G. Borghs, I.I. Kaya, and
R.A. Stradling, Phys.~Rev.~B {\bf 63}, 052504 (2001); 
A.N.~Grigorenko, S.J.~Bending, M.J.~Van Bael, M.~Lange, V.V.~Moshchalkov, 
H.~Fangohr, and P.A.J.~de Groot, 
Phys.~Rev.~Lett.~{\bf 90}, 237001 (2003).   

\bibitem{Jensen}
C.~Reichhardt and N. Gr{\o}nbech-Jensen, 
Phys.~Rev. Lett. {\bf 85}, 2372 (2000).  

\bibitem{Schuller}
C.~Reichhardt, G.T.~Zim{\' a}nyi, R.T.~Scalettar, 
A.~Hoffmann, and I.K.~Schuller, Phys.~Rev.~B {\bf 64}, 052503 (2001).  

\bibitem{Peeters}
G.R.~Berdiyorov, M.V.~Milosevic, and F.M.~Peeters, 
Phys.~Rev.~Lett. {\bf 96}, 207001 (2006); 
Phys.~Rev.~B {\bf 74}, 174512 (2006).   

\bibitem{Reichhardt}
C.~Reichhardt, C.J. Olson, and F.~Nori, Phys.~Rev.~B {\bf 57}, 7937 (1998).  

\bibitem{Chen} 
Q.H.~Chen, G.~Teniers, B.B.~Jin and V.V.~Moshchalkov, 
Phys.~Rev.~B {\bf 73}, 014506 (2006). 

\bibitem{Milosvic}
G.R.~Berdiyorov, M.V.~Milosevic, and F.M.~Peeters,
Europhys.~Lett.~{\bf 74}, 493 (2006). 

\bibitem{Zimanyi1}
C.~Reichhardt, C.J.~Olson, R.T.~Scalettar, 
and G.T.~Zim{\' a}nyi, Phys.~Rev.~B {\bf 64}, 144509 (2001).  

\bibitem{Olson}
C.J. Olson Reichhardt, A.~Lib{\' a}l, and C.~Reichhardt, 
Phys.~Rev.~B {\bf 73}, 184519 (2006). 

\bibitem{Zimanyi}
C. Reichhardt, G.T.~Zim{\' a}nyi, and N. Gr{\o}nbech-Jensen,
Phys.~Rev.~B {\bf 64}, 014501 (2001). 

\bibitem{Molecular}
C.~Reichhardt and C.J.~Olson Reichahrdt, 
Phys.~Rev.~B {\bf 76}, 064523 (2007).

\bibitem{Dom}
M.F.~Laguna, C.A.~Balseiro, D.~Dom{\' \i}nguez, and F.~Nori,
Phys.~Rev.~B {\bf 64}, 104505 (2001). 

\bibitem{Peet}
G.R.~Berdiyorov, V.R.~Misko, M.V.~Milosevic, W.~Escoffier, 
I.V.~Grigorieva, and F.M.~Peeters,
Phys.~Rev.~B {\bf 77}, 024526 (2008). 

\bibitem{Shapirolook}
L. Van Look, E. Rosseel, M.J. Van Bael, K. Temst, V.V. Moshchalkov,
and Y. Bruynseraede,
Phys. Rev. B {\bf 60}, R6998 (1999).

\bibitem{Wu}
T.C.~Wu, P.C.~Kang, L.~Horng, J.C.~Wu, and T.J.~Yang,
J.~Appl.~Phys.~{\bf 95}, 6696 (2004).  

\bibitem{Cao}
R.~Cao, T.C.~Wu, P.C.~Kang, J.C.~Wu, T.J.~Yang, and L.~Horng,
Sol. St.~Commun. {\bf 143}, 171 (2007).  

\bibitem{Look2}
L.~Van Look, B.Y.~Zhu, R.~Jonckheere, B.R.~Zhao, Z.X.~Zhao, and 
V.V.~Moshchalkov,
Phys.~Rev.~B {\bf 66}, 214511 (2002). 

\bibitem{Velez}
M.~Velez, D.~Jaque, J.I.~Mart{\' \i}n, M.I. Montero, 
I.K.~Schuller, and J.L.~Vicent,
Phys.~Rev.~B {\bf 65}, 104511 (2002).  

\bibitem{Villegas2}
J.E.~Villegas, E.M.~Gonzalez, M.I.~Montero, I.K.~Schuller and J.L.~Vicent,
Phys.~Rev.~B {\bf 72}, 064507 (2005).  

\bibitem{Bechinger}
K.~Mangold, P.~Leiderer, and C.~Bechinger, 
Phys.~Rev.~Lett.~{\bf 90}, 158302 (2003). 

\bibitem{Tierno}
P.~Tierno, T.H.~Johansen, and T.M.~Fischer,
Phys.~Rev.~Lett.~{\bf 99}, 038303 (2007).  

\bibitem{Coupier}
G.~Coupier, M. Saint Jean and C.~Guthmann, Phys.~Rev.~B {\bf 75}, 224103 (2007). 
\bibitem{Zimanyi2}
C. Reichhardt, K.~Moon, R.~Scalettar, and G.~Zim{\' a}nyi,
Phys.~Rev.~Lett.~{\bf 83}, 2282 (1999). 

\bibitem{Fertig}
M.-C. Cha and H.A. Fertig, Phys.~Rev.~Lett.~{\bf 80}, 3851 (1998). 

\bibitem{Jensen2}
C. Reichhardt and N. Gr{\o}nbech-Jensen,
Phys.~Rev.~Lett. {\bf 85}, 2372 (2000).

\bibitem{Reichhardt2001}
C. Reichhardt and N. Gr{\o}nbech-Jensen,
Phys.~Rev.~B {\bf 63}, 054510 (2001).


\end{thebibliography}
\end{document}